\documentclass[aps,prd,preprint,nofootinbib,11pt]{revtex4}
\usepackage{amsfonts}
\usepackage{mathrsfs}
\usepackage{graphicx}
\usepackage{amsmath}
\usepackage{amssymb}
\usepackage{multirow}
\usepackage{subfigure}
\usepackage{epsfig}
\usepackage{graphicx}
\usepackage{color}
\newcommand{\bqa}{\begin{eqnarray}}
\newcommand{\eqa}{\end{eqnarray}}
\newcommand{\beq}{\begin{equation}}
\newcommand{\eeq}{\end{equation}}
\allowdisplaybreaks[1]
\graphicspath{{fig/}{dia/}} \DeclareGraphicsExtensions{.eps}
\begin{document}

\title{Spectrum of $[8]_{[c\bar{s}]} \otimes [8]_{[q \bar{q^\prime}]}$ systems with quantum numbers $J^{P}=0^\pm$ and $1^\pm$\\[0.7cm]}
\author{Si-Nuo Li$^{1}$ and Liang Tang$^{1}$\footnote{tangl@hebtu.edu.cn}}

\affiliation{$^1$ College of Physics and Hebei Key Laboratory of Photophysics Research and Application, Hebei Normal University, Shijiazhuang 050024, China}



\begin{abstract}
\vspace{0.3cm}
Inspired by the recent experimental progress on the $T_{c\bar{s}0}^a(2900)^{0/++}$, the fully open tetraquark spectrum with the configuration of $[8]_{[c\bar{s}]} \otimes [8]_{[q \bar{q^\prime}]}$ is systematically investigated by the QCD sum rules. In this article, we concentrate on the quantum numbers $J^{P}=0^{+}/0^{-}/1^{+}/1^{-}$.

Firstly, we construct four scalar currents ($J^{P}=0^+$) in the form of $[8]_{[c\bar{s}]} \otimes [8]_{[u\bar{d}]}$ type tetraquark structure and perform the analysis of the QCD sum rules, where we consider the leading order contributions up to dimension 11 in the operator product expansion and retain contributions linear in the strange quark mass $m_s$. Our results, $M_{0^{+}}^A = 2.91_{-0.20}^{+0.20}~\text{GeV}$ and $M_{0^{+}}^C = 2.98_{-0.21}^{+0.20}~\text{GeV}$, are consistent with the experimentally discovered $T_{c\bar{s}0}^a(2900)^{++}$ within error margins. Thus, our calculation supports classifying $T_{c\bar{s}0}^a(2900)^{++}$ as a tetraquark state with the $[8]_{[c\bar{s}]} \otimes [8]_{u \bar{d}}$ color configuration.

Moreover, on the basis of $J^P=0^+$, we also study tetraquark states with quantum numbers $J^P=0^-$, $1^+$, and $1^-$, predicting two new hadronic states awaiting experimental verification for each quantum number. Their masses are as follows: $M_{0^{-}}^{A}=3.45_{-0.10}^{+0.12}~\text{GeV}$, $M_{0^{-}}^{D}=3.20_{-0.19}^{+0.17}~\text{GeV}$, $M_{1^{+}}^{A}=2.95_{-0.19}^{+0.19}~\text{GeV}$, $M_{1^{+}}^{C}={2.95_{-0.19}^{+0.19}}~\text{GeV}$, $M_{1^{-}}^{A}={3.23_{-0.10}^{+0.11}}~\text{GeV}$ and $M_{1^{-}}^{D}={3.23_{-0.11}^{+0.09}}~\text{GeV}$. With advances in experimental techniques and accumulation of new data, these predicted results are hoped to be confirmed in future experiments.
\end{abstract}
 \maketitle
\newpage

\section{Introduction}
For a long time, it was believed that hadrons were classifiable into two groups: mesons, comprising a quark and an antiquark ($q \bar{q}$), and baryons, made up of three quarks ($qqq$)~\cite{Gell-Mann:1964ewy,Zweig:1964ruk}. Nonetheless, quantum chromodynamics (QCD), the foundational theory of strong interactions, theoretically permits the existence of any color-singlet configurations such as tetraquarks, pentaquarks, hybrids, and glueballs among others. Over the past two decades, numerous potential exotic states have been experimentally reported, beginning with the Belle Collaboration's discovery of the X(3872) in 2003~\cite{CDF:2003cab}. These candidates have sparked significant interest in theoretical research, leading to a substantial number of publications investigating the nature and characteristics of these potential exotic states. For comprehensive reviews on the current experimental and theoretical research into exotic states, refer to References~\cite{Ali:2017jda, Guo:2017jvc, Esposito:2016noz, Lebed:2016hpi, Richard:2016eis, Chen:2016qju, Liu:2019zoy, Brambilla:2019esw}.

In 2020, the LHCb Collaboration announced the discovery of two isosinglet exotic states, $X_0(2900)$ and $X_1(2900)$, observed in the $D^-K^+$ invariant mass spectrum~\cite{LHCb:2020pxc}. These states are naturally characterized by a $\bar{c}\bar{s}ud$ flavor structure. Recently, in 2022, a doubly charged tetraquark and its neural partner were reported by the LHCb Collaboration~\cite{LHCb:2022sfr}, which have spin-parity $J^P=0^+$ and quark constituent $[c\bar{s} u \bar{d}]$ and $[c\bar{s} d \bar{u}]$, respectively. The masses of these two new resonant states are $2.908\pm0.011\pm0.020$ GeV, and widths are $0.136\pm0.023\pm0.013$ GeV. Therefore, they are nominated as $T_{c\bar{s}0}^a(2900)^{++/0}$. From the perspective of isospin analysis, they should be two components of an isospin vector, namely $T_{c\bar{s}0}^a(2900)^{0}$ and $T_{c\bar{s}0}^a(2900)^{++}$ are the components with $I_{3} = -1$ and $1$, respectively.

These states are composed of four distinct flavor quarks (or antiquarks), setting them apart from exotic states confirmed previously. The $X_{0/1}(2900)$ particle represents the inaugural discovery of an exotic structure with entirely open flavor, sparking numerous research endeavors aimed at unraveling its internal structure, including the tetraquark state~\cite{Lu:2020qmp, Wang:2020xyc, He:2020jna, Karliner:2020vsi, Zhang:2020oze, Agaev:2021jsz}, molecular state~\cite{Hu:2020mxp, Liu:2020nil, Chen:2020aos, Huang:2020ptc}, triangle singularity~\cite{Liu:2020orv}, and cusp effect~\cite{Burns:2020epm}. The fully open flavor tetraquark states $T_{c\bar{s}0}^a(2900)^{++/0}$ could potentially be the isospin counterparts of the tetraquark candidates made up of $[c\bar{s}u\bar{d}]$ and $[c\bar{s}d\bar{u}]$, thereby garnering significant research interest aimed at understanding their inner structures. Very recently, the $T_{c\bar{s}0}^a(2900)^{++/0}$ have been investigated by various theoretical methods, containing the potential quark model~\cite{Liu:2022hbk}, the color flux-tube model~\cite{Wei:2022wtr}, the one-boson-exchange model~\cite{Chen:2022svh}, the effective Lagrantian approach~\cite{Yue:2022mnf}, and the framework of the QCD sum rules~\cite{Agaev:2022duz, Agaev:2022eyk, Yang:2023evp}. The authors in Ref.~\cite{Wei:2022wtr} carried out a systematical investigation on the properties of the ground and P-wave states $[cs][\bar{u} \bar{d}]$ and $[cu][\bar{s}\bar{d}]$ with various spin, isospin or U-spin, and color combinations in a multiquark color flux-tube model. In Ref.~\cite{Chen:2022svh}, based on the one-boson-exchange model to study the $D^{(*)}K^{*}$ interactions, Chen and Huang could simultaneously reproduced the masses of the $T_{c\bar{s}0}^a(2900)^{0/++}$ in the hadronic molecular picture, where the $T_{c\bar{s}0}^a(2900)^{0/++}$ are regarded as the $D^{*}K^{*}[1(0^+)]$ charmed-strange molecular states. In Ref.~\cite{Agaev:2022duz}, Agaev $\textit{et al}.$  investigated the new scalar resonances $T_{c\bar{s}0}^a(2900)^{++}$ and $T_{c \bar{s}0}^a(2900)^{0}$ as molecules $D_s^{*+} \rho^+$ and $D_{s}^{*+} \rho^-$ of conventional vector mesons, respectively, within QCD sum rules. Basing on the analysis of the QCD sum rules~\cite{Yang:2023evp}, Yang $\textit{et al.}$ assigned the $T_{c\bar{s}0}^a(2900)^{++/0}$ as the $A-\bar{A}$-type tetraquark states in the framework of the QCD sum rules, where the A denotes the axialvector diquark state.

It should be highlighted that QCD suggests the presence of a tetraquark configuration, consisting of two color-octet components. This configuration, due to QCD interactions, distinguishes itself from a molecular state comprised of two color-singlet mesons, allowing for the possibility of decay into two mesons via the exchange of one or more gluons. Thus, exploring the color octet-octet tetraquark state holds significant importance for identifying potential new hadronic states \cite{Wang:2006ri, Yang:2020wkh, Wang:2021mma}. In this work, we systematically construct the color octet-octet tetraquark currents comprised of $[8]_{c\bar{s}} \otimes [8]_{q\bar{q^\prime}}$ ($[q \bar{q^\prime}] = [u\bar{d}]$ or $[d\bar{u}]$) with quantum numbers $J^P= 0^{\pm}$ and $1^{\pm}$ and determine their masses using QCD sum rules. Subsequently, we analysis the possibility of $T_{c\bar{s}0}^a(2900)^{0/++}$ as color octet-octet states and discuss the predictions with $0^-$ and $1^{\pm}$.

Following the introduction, Sect.~\ref{sec:formalism} outlines the fundamental equations of the QCD sum rules. Sect.~\ref{sec:numerical} is dedicated to numerical analyses and the presentation of corresponding figures. Finally, we provide the conclusion and discussion of the tetraquark states with fully open flavors in Sec.~\ref{sec:discussion}.

\section{Formalism}\label{sec:formalism}
The research content of this work involves the study of the color octet-octet tetraquark states with quantum numbers $J^{P} = 0^{\pm}$ and $1^{\pm}$ using the QCD sum rule method. The starting point of the QCD sum rules is the two-point correlation function constructed from two hadronic currents. For the scalar states considered in this paper, the form of the two-point correlation function is as follows:
\begin{eqnarray}
\Pi(q^{2}) &=& i \int d^4 x e^{i q \cdot x}\langle 0 | T \{j(x), j^{\dagger}(0)\} |0 \rangle .
\end{eqnarray}

Here, $j(x)$and $j(0)$ are hadronic currents with the aforementioned scalar quantum number. By consulting literature and conducting independent analyses, we construct the following four types of color octet-octet structure scalar currents:
\begin{eqnarray}
j_{A}^{0^+}(x) &=& [ i \bar{s}^{j} (x) \gamma^{5} (t^{a})_{jk} c^{k}(x) ][ i \bar{d}^{m} (x) \gamma^{5} (t^{a})_{mn} u^{n}(x) ], \label{current-A}\\
j_{B}^{0^+}(x) &=& [ \bar{s}^{j} (x) (t^{a})_{jk} c^{k}(x) ][ \bar{d}^{m} (x) (t^{a})_{mn} u^{n}(x) ], \label{current-B}\\
j_{C}^{0^+}(x) &=& [ \bar{s}^{j} (x) \gamma^{{\mu}} (t^{a})_{jk} c^{k}(x) ][ \bar{d}^{m} (x) \gamma_{{\mu}} (t^{a})_{mn} u^{n}(x) ],\label{current-C} \\
j_{D}^{0^+}(x) &=& [ \bar{s}^{j} (x) \gamma^{{\mu}} \gamma^{5} (t^{a})_{jk} c^{k}(x) ][ \bar{d}^{m} (x) \gamma_{{\mu}} \gamma^{5} (t^{a})_{mn} u^{n}(x) ]. \label{current-D}
\end{eqnarray}
Here, $j$, $k$, $m$, and $n$ are color indices, $t^a = \frac{\lambda^a}{2}$, where $\lambda^a$ are the Gell-Mann matrices, and $t^a$ are the generators of the $SU(3)_c$ group; the subscripts $A$, $B$, $C$, and $D$ respectively represent currents composed of two $0^{-}$, two $0^{+}$, two $1^{-}$, and two $1^{+}$ color-octet components. In the construction process, by swapping $u \leftrightarrow d$, we can obtain the corresponding currents within the same isospin multiplet. In the isospin limit, the tetraquark states within the same isospin multiplet have the same mass. It is noteworthy that between the two color-octet components within each current, there is an S-wave interaction, which is the reason we only study these four currents in this paper.

Similarly, we present the $[8]_{c\bar{s}}\otimes [8]_{u\bar{d}}$ currents for quantum numbers $0^-$ and $1^{\pm}$ below, respectively:
\begin{eqnarray}
j_{A}^{0^{-}}(x) &=& [ i \bar{s}^{j} (x) \gamma^{5} (t^{a})_{jk} c^{k}(x) ][ i \bar{d}^{m} (x) (t^{a})_{mn} u^{n}(x) ],\label{current0-A}\\
j_{B}^{0^{-}}(x) &=& [ \bar{s}^{j} (x) (t^{a})_{jk} c^{k}(x) ][ \bar{d}^{m} (x) \gamma^{5} (t^{a})_{mn} u^{n}(x) ],\label{current0-B}\\
j_{C}^{0^{-}}(x) &=& [ \bar{s}^{j} (x) \gamma^{{\mu}} (t^{a})_{jk} c^{k}(x) ][ \bar{d}^{m} (x)\gamma_{{\mu} }\gamma^{5} (t^{a})_{mn} u^{n}(x) ],\label{current0-C}\\
j_{D}^{0^{-}}(x) &=& [ \bar{s}^{j} (x) \gamma^{{\mu}} \gamma^{5} (t^{a})_{jk} c^{k}(x) ][ \bar{d}^{m} (x) \gamma_{\mu} (t^{a})_{mn} u^{n}(x) ]\label{current0-D},
\end{eqnarray}
\begin{eqnarray}
j_{A}^{1^{+}}(x) &=& [ i \bar{s}^{j} (x) \gamma^{5} (t^{a})_{jk} c^{k}(x) ][ \bar{d}^{m} (x) \gamma_{{\mu}} (t^{a})_{mn} u^{n}(x) ] , \label{current-1+A}\\
j_{B}^{1^{+}}(x) &=& [ \bar{s}^{j} (x) (t^{a})_{jk} c^{k}(x) ][ \bar{d}^{m} (x) \gamma_{{\mu}} \gamma^{5} (t^{a})_{mn} u^{n}(x) ],\label{current-1+B}\\
j_{C}^{1^{+}}(x) &=& [ \bar{s}^{j} (x) \gamma^{{\mu}} (t^{a})_{jk} c^{k}(x) ][ i \bar{d}^{m} (x) \gamma^{5} (t^{a})_{mn} u^{n}(x) ],\label{current-1+C}\\
j_{D}^{1^{+}}(x) &=& [ \bar{s}^{j} (x) \gamma^{{\mu}} \gamma^{5} (t^{a})_{jk} c^{k}(x) ][ \bar{d}^{m} (x) (t^{a})_{mn} u^{n}(x) ]\label{current-1+D},
\end{eqnarray}
\begin{eqnarray}
j_{A}^{1^{-}}(x) &=& [ i \bar{s}^{j} (x) \gamma^{5} (t^{a})_{jk} c^{k}(x) ][ \bar{d}^{m} (x) \gamma_{{\mu}} \gamma^{5} (t^{a})_{mn} u^{n}(x) ],\label{current-1-A}\\
j_{B}^{1^{-}}(x) &=& [ \bar{s}^{j} (x) (t^{a})_{jk} c^{k}(x) ][ \bar{d}^{m} (x) \gamma_{{\mu}} (t^{a})_{mn} u^{n}(x) ],\label{current-1-B}\\
j_{C}^{1^{-}}(x) &=& [ \bar{s}^{j} (x) \gamma^{{\mu}} (t^{a})_{jk} c^{k}(x) ][ \bar{d}^{m} (x) (t^{a})_{mn} u^{n}(x) ],\label{current-1-C}\\
j_{D}^{1^{-}}(x) &=& [ \bar{s}^{j} (x) \gamma^{{\mu}} \gamma^{5} (t^{a})_{jk} c^{k}(x) ][ i \bar{d}^{m} (x) \gamma^{5} (t^{a})_{mn} u^{n}(x) ]\label{current-1-D}.
\end{eqnarray}

The quark-hadron duality principle is a fundamental assumption of the QCD sum rule method: on one hand, the correlation function $\Pi(q^2 )$ can be computed at the quark-gluon level using the Operator Product Expansion (OPE); on the other hand, it can be represented at the hadron level by introducing hadronic coupling constant and mass; the two representations are equivalent under an appropriate energy scale, which is the essence of the quark-hadron duality principle. For convenience, we refer to the calculation at the OPE level as the OPE side of the correlation function, and the representation at the hadron level as the phenomenological side of the correlation function.

To calculate the spectral density on the OPE side, we utilize the full propagators for heavy quarks ($Q=c,b$) and light quarks ($q=u,d,s$), denoted as $S_{ij}^Q(p)$ and $S_{ij}^q(x)$, respectively:
\begin{eqnarray}
S^Q_{ij}(p)\!\!\!&=&\!\!\!\frac{i \delta_{ij}(p\!\!\!\slash + m_Q)}{p^2 - m_Q^2} - \frac{i}{4} \frac{t^a_{ij} G^a_{\alpha\beta} }{(p^2 - m_Q^2)^2} [\sigma^{\alpha \beta}
(p\!\!\!\slash + m_Q)
+ (p\!\!\!\slash + m_Q) \sigma^{\alpha \beta}] \nonumber \\\!\!\!&+&\!\!\!  \frac{i\delta_{ij}m_Q  \langle g_s^2 G^2\rangle}{12(p^2 - m_Q^2)^3}\bigg[ 1 + \frac{m_Q (p\!\!\!\slash + m_Q)}{p^2 - m_Q^2} \bigg] \nonumber \\\!\!\!&+&\!\!\!\frac{i \delta_{ij}}{48} \bigg\{ \frac{(p\!\!\!\slash +
m_Q) [p\!\!\!\slash (p^2 - 3 m_Q^2) + 2 m_Q (2 p^2 - m_Q^2)] }{(p^2 - m_Q^2)^6}
\times (p\!\!\!\slash + m_Q)\bigg\} \langle g_s^3 G^3 \rangle \; ,\label{SQ}
\end{eqnarray}
\begin{eqnarray}
S_{ij}^{q}(x)\!\!\!&=&\!\!\!\frac{i \delta_{ij} x\!\!\!\slash }{2 \pi^{2} x^{4}}-\frac{m_{q} \delta_{ij}}{4 \pi^{2} x^{2}}-\frac{i t^{N}_{ij}}{32 \pi^{2} x^{2}} (\sigma_{\alpha \beta} x\!\!\!\slash +x\!\!\!\slash \sigma_{\alpha \beta})\nonumber\\&-&\frac{\delta_{ij}}{12} \langle \bar{q} q \rangle +\frac{i \delta_{ij} x\!\!\!\slash}{48} m_{q}\langle \bar{q} q \rangle -\frac{\delta_{ij} x^{2}}{192}  \langle \bar{q} G q \rangle \nonumber\\&+&
\frac{i \delta_{ij} x^{2} x\!\!\!\slash}{1152} m_{q} \langle \bar{q} G q \rangle -\frac{t^{N}_{ij} \sigma_{\alpha \beta}}{192} \langle \bar{q} G q \rangle \nonumber\\&+&
\frac{i t^{N}_{ij} }{768} (\sigma_{\alpha \beta} x\!\!\!\slash +x\!\!\!\slash \sigma_{\alpha \beta}) m_{q} \langle \bar{q} G q \rangle\;,\label{Sq}
\end{eqnarray}
where the vacuum condensates of QCD operators are clearly displayed. The complete propagators for both light and heavy quarks have been extensively derived and explained in the literature~\cite{Reinders:1984sr}. Therefore, once we obtain the spectral density $\rho_i^{\text{OPE}}(s)$ using the full propagators, we can determine the correlation function $\Pi(q^{2})$ at the quark-gluon level based on the dispersion relation:
\begin{eqnarray}
  \Pi^{\text{OPE}}_{i} (q^2) = \int_{(m_s +m_c)^2}^{s_0} ds \frac{\rho_{i}^{\text{OPE}}(s)}{s - q^2} + \Pi_{i}^{\text{cond}}(q^2), \label{Pi-OPE}
\end{eqnarray}
where, the subscript $i$ ranges from $A$ to $D$, with the first term $\rho_{i}^{\text{OPE}}(s) = \text{Im} [\Pi_{i}^{\text{OPE}}(s)]/\pi$, and the second term $\Pi_i^{\text{cond}}(q^2)$ represents the terms that can be directly transformed by the Borel transformation without the need for a dispersion relation. $\rho_{i}^{\text{OPE}}(s)$ can be expanded as follows:
\begin{eqnarray}
\rho^{\text{OPE}}_{i}(s) &=& \rho^{\text{pert}}_{i}(s) + \rho^{\langle \bar{s} s \rangle}_{i}(s) +\rho^{\langle G^{2} \rangle}_{i}(s) + \rho^{\langle \bar{s} G s\rangle} + \rho^{\langle G^{3} \rangle}_{i}(s) + \rho_{i}^{\langle \bar{q} q\rangle^{2}}(s) \nonumber \\ &+& \rho_{i}^{\langle \bar{s} G s\rangle}(s) + \rho_{i}^{\langle \bar{s} s\rangle \langle G^{2} \rangle}(s) + \rho_{i}^{\langle \bar{s} s\rangle \langle G^{2} \rangle}(s) +\rho_{i}^{\langle \bar{q} q\rangle \langle \bar{q} G  q\rangle}(s) + \rho_{i}^{\langle G^{2} \rangle^{2}}(s).
\label{rho-OPE}
\end{eqnarray}

Applying the Borel transformation to the OPE side, namely equation (\ref{Pi-OPE}), we have:
\begin{eqnarray}
  \Pi^{\text{OPE}}_i (M^{2}_{B}) &=& \int_{(m_s + m_c)^2}^{\infty} ds \rho^{\text{OPE}}(s)e^{-s/M_B^2} + \Pi_{i}^{\langle G^{3} \rangle}(M^{2}_{B}) + \Pi_{i}^{\langle G^{2} \rangle \langle \bar{s} s \rangle}(M^{2}_{B}) \nonumber \\
  & &+ \Pi_{i}^{\langle \bar{s} s \rangle \langle \bar{q} G q \rangle }(M^{2}_{B}) + \cdots.
  \label{Pi-OPE-MBMB}
\end{eqnarray}

To account for the effect induced by the mass of the strange quark, we retain the linear terms of the strange quark mass in our subsequent calculations. Taking the current defined by equation (\ref{current-A}) as an example, the spectral density obtained through analytical calculation is presented as follows:
\begin{eqnarray}
\rho_{A}^{\text{pert}}(s) &=& -\frac{1}{ 18432 \pi^{6}} \int^{\lambda}_{0} d x \frac{H_{x}^{3} (H_{x} - 4 m_{c} m_{s} x )}{ (x-1)^{3} x}, \\
\rho_{A}^{\langle \bar{s} s\rangle}(s) &=& - \frac{\langle \bar{s} s\rangle}{384 \pi^{4} } \int^{\lambda}_{0} d x \frac{H_{x}^{2} ( m_{s} (x-1)+m_{c} x )}{(x-1)^{2} x},\\
\rho_{A}^{\langle G^{2} \rangle}(s) &=& \frac{\langle g_{s}^{2} G^{2} \rangle}{55296 \pi^{6} } \int^{\lambda}_{0} d x \frac{ m_{c} x^{2} ( -H_{x} m_{c} + 3 H_{x} m_{s} + m_{c}^{2} m_{s} x )}{(x-1)^{3}} \nonumber \\ & &-\frac{\langle g_{s}^{2} G^{2} \rangle}{98304 \pi^{6} } \int^{\lambda}_{0} d x  \frac{ H_{x}^{2}}{ (x-1)^{2}} +
\frac{\langle g_{s}^{2} G^{2} \rangle}{49152 \pi^{6}} \int^{\lambda}_{0} d x \frac{ H_{x} (H_{x} - 2 m_{c} m_{s})}{ (x-1)x},\\
\rho_{A}^{\langle \bar{s} G s\rangle}(s) &=& \frac{\langle \bar{s} g_{s} \sigma \cdot G  s \rangle}{1152 \pi^{4} } \int^{\lambda}_{0} d x  \frac{ H_{x} ( m_{s} (x-1) + 3 m_{c} x )}{(x-1) x} \\ \nonumber
 &-&  \frac{\langle \bar{s} g_{s} \sigma \cdot G s\rangle}{3072 \pi^{4} } \int^{\lambda}_{0} d x \frac{H_{x} (m_{s} (x-1)+ m_{c} x)}{(x-1)^{2}},
\end{eqnarray}
\begin{eqnarray}
\rho_{A}^{\langle G^{3} \rangle}(s) &=& \frac{\langle g_{s}^{3} G^{3} \rangle}{221184 \pi^{6}} \int^{\lambda}_{0} d x \frac{ x^{2} ( H_{x} +2 m_{c} ( m_{c} - 3 m_{s}) x )}{ (x-1)^{3}} \\ \nonumber
 &+&\frac{\langle g_{s}^{3} G^{3} \rangle}{589824 \pi^{6}} \int^{\lambda}_{0} d x  \frac{ x ( 3 H_{x} +2 m_{c}^{2} x)}{ (x-1)^{2}},\\
\rho_{A}^{\langle \bar{q} q\rangle^{2}}(s) &=& - \frac{\langle \bar{q} q\rangle^{2}}{72 \pi^{2} } \int^{\lambda}_{0} d x \frac{-H_{x}+m_{c} m_{s} x }{x}, \\
\rho_{A}^{\langle \bar{s} s\rangle \langle G^{2} \rangle}(s) &=& - \frac{\langle \bar{s} s\rangle \langle g_{s}^{2} G^{2} \rangle}{4608 \pi^{4} } \int^{\lambda}_{0} d x \frac{ m_{c} x^{2} }{(x-1)^{2}} \\ \nonumber
&-& \frac{\langle \bar{s} s\rangle \langle g_{s}^{2} G^{2} \rangle}{36864 \pi^{4} } (m_{c}- 2 m_{s}),\\
\rho_{A}^{\langle \bar{q} q\rangle \langle \bar{q} G  q\rangle}(s) &=& - \frac{\langle \bar{q} q\rangle \langle \bar{q} g_{s} \sigma \cdot G q\rangle}{144 \pi^{2} },\\
\rho_{A}^{\langle G^{2} \rangle^{2}}(s) &=& \frac{\langle g_{s}^{2} G^{2}\rangle^{2}}{10616832 \pi^{6} } \int^{\lambda}_{0} d x \frac{H_{x}^{3}(2 - 5x )+ 6 x^{4} }{(x-1)x^{3}} + \frac{\langle g_{s}^{2} G^{2}\rangle^{2}}{196608 \pi^{6} },
\end{eqnarray}
where $M_B^2$ is the Borel parameter, $H_x = m_c^2 x - x (1-x)s$, and $\lambda = 1-m_c^2/s$.

The results of the direct Borel transformation applied to equation (\ref{Pi-OPE}), without going through the dispersion relation, are as follows:
\begin{eqnarray}
\Pi_{A}^{\langle G^{3} \rangle}(M^{2}_{B}) &=& \frac{\langle g_{s}^{3} G^{3}\rangle}{110592 \pi^{6} } \int^{1}_{0} d x \frac{m_{c}^{3} m_{s} x^{3}}{(x-1)}e^{\frac{m_{c}^{2}}{ (x-1) M_{B}^{2}}},\\
\Pi_{A}^{\langle G^{2} \rangle \langle \bar{s} s \rangle}(M^{2}_{B}) &=& -\frac{\langle g_{s}^{2} G^{2}\rangle \langle \bar{s} s \rangle}{13824 \pi^{4} } \int^{1}_{0} d x \frac{m_{c}^{2} x (m_{s} ( x - 1) + m_{c} x) }{(x-1)^{3}}e^{\frac{m_{c}^{2}}{ (x-1) M_{B}^{2}}},\\
\Pi_{A}^{\langle \bar{s} s \rangle \langle \bar{q} G q \rangle }(M^{2}_{B}) &=& -\frac{\langle \bar{s} s \rangle \langle \bar{q} g_{s} \sigma \cdot G q \rangle}{144 \pi^{2} } \int^{1}_{0} d x m_{c} m_{s} e^{\frac{m_{c}^{2}}{ (x-1) M_{B}^{2}}},\\
\Pi_{A}^{\langle G^{2} \rangle^{2}}(M^{2}_{B}) &=& \frac{\langle g_{s}^{2} G^{2} \rangle ^{2}}{10616832 \pi^{6} } \int^{1}_{0} d x \{ \frac{-8 m_{c}^{3} m_{s} x }{M_{B}^{2} (x-1)^{3}}\nonumber \\
& & +\frac{M_{B}^{2} m_{c}(x-1)(8 m_{c} x - m_{s}(2+21x))}{M_{B}^{2} (x-1)^{3}}\}  e^{\frac{m_{c}^{2}}{ (x-1) M_{B}^{2}}},\\
\Pi_{A}^{\langle G^{3} \rangle \langle \bar{s} s \rangle }(M^{2}_{B}) &=&- \frac{\langle g_{s}^{3} G^{3} \rangle \langle \bar{s} s \rangle}{55296 \pi^{4} } \int^{1}_{0} d x  \{ 2 m_{c}^{2}\frac{m_{s} (x-1)+m_{c} x}{M_{B}^{2} (x-1)^{4}}\nonumber \\
& &+2 m_{c}^{2}\frac{M_{B}^{2} (x-1)m_{s} (x-1)}{M_{B}^{2} (x-1)^{4}}\} e^{\frac{m_{c}^{2}}{ (x-1) M_{B}^{2}}},\\
\Pi_{A}^{\langle G^{2} \rangle \langle \bar{s} G s \rangle }(M^{2}_{B}) &=& \frac{\langle g_{s}^{2} G^{2} \rangle \langle \bar{q} g_{s} \sigma \cdot G q\rangle}{82944 \pi^{4} } \int^{1}_{0} d x \frac{m_{c}
 (m_{s} m_{c} (x-1)) + 3m_{c}^{2} x +9M_{B}^{2}(x-1)x}{M_{B}^{2} (x-1)^{3}}  e^{\frac{m_{c}^{2}}{ (x-1) M_{B}^{2}}}\nonumber \\ &+&\frac{\langle g_{s}^{2} G^{2} \rangle \langle \bar{q} g_{s} \sigma \cdot G q\rangle}{221184 \pi^{4} } \int^{1}_{0} d x \frac{m_{c}^{2} m_{s}+ M_{B}^{2} (-3 m_{c} + m_{s})}{M_{B}^{2}}  e^{\frac{m_{c}^{2}}{ (x-1) M_{B}^{2}}},\\
\Pi_{A}^{\langle G^{3} \rangle \langle G^{2} \rangle}(M^{2}_{B}) &=& \frac{\langle g_{s}^{3} G^{3} \rangle \langle g_{s}^{2} G^{2} \rangle}{7077888 \pi^{6} } \int^{1}_{0} d x \{ \frac{(-2 m_{c}^{3} m_{s}x +2M_{B}^{2} m_{c} x(-3 m_{s} + m_{c})(x-1)}{M_{B}^{2} (x-1)^{4}}\nonumber \\ & &
\frac{m_{c}(x-1)+ M_{B}^{2} (x-1)^{2}) x }{M_{B}^{2} (x-1)^{4}} \}  e^{\frac{m_{c}^{2}}{ (x-1) M_{B}^{2}}},\\
\Pi_{A}^{\langle \bar{q} G q \rangle^{2} }(M^{2}_{B}) &=& \frac{\langle \bar{q} g_{s} \sigma \cdot G q\rangle^{2}}{1152 \pi^{2} } \int^{1}_{0} d x \frac{M_{B}^{4} + M_{B}^{2} m_{c}^{2} + m_{c}^{3} m_{s}}{M_{B}^{2} } e^{\frac{m_{c}^{2}}{ (x-1) M_{B}^{2}}},\\
\Pi_{A}^{\langle G^{2} \rangle \langle \bar{q} q \rangle^{2}}(M^{2}_{B}) &=&- \frac{\langle g_{s}^{2} G^{2} \rangle \langle \bar{q} q \rangle^{2}}{5184 \pi^{2} } \int^{1}_{0} d x \frac{m_{c} (-m_{c}^{3} m_{s} + M_{B}^{4} m_{c} (m_{c} - 3m_{s})(x-1))}{M_{B}^{2} (x-1)^{3}}  e^{\frac{m_{c}^{2}}{ (x-1) M_{B}^{2}}}\nonumber \\ & &- \frac{\langle g_{s}^{2} G^{2} \rangle \langle \bar{q} q \rangle^{2}}{13824 \pi^{2} } \int^{1}_{0} d x   e^{\frac{m_{c}^{2}}{ (x-1) M_{B}^{2}}},\\
\Pi_{A}^{\langle G^{3} \rangle \langle \bar{s} G s\rangle}(M^{2}_{B}) &=& \frac{\langle g_{s}^{3} G^{3} \rangle \langle \bar{s} g_{s} \sigma \cdot G s\rangle}{331776 \pi^{4} } \int^{1}_{0}d x \{ \frac{2 m_{c}^{2} (m_{s} (x-1)) +3 m_{c} x}{M_{B}^{4} (x-1)^{4}} \nonumber \\
& &+ \frac{M_{B}^{2} (x-1) (m_{s} (x-1) +18 m_{c} x)}{M_{B}^{4} (x-1)^{4}}\} e^{\frac{m_{c}^{2}}{ (x-1) M_{B}^{2}}},\\
\Pi_{A}^{\langle \bar{s} s \rangle \langle \bar{q} q \rangle \langle \bar{q} G q\rangle}(M^{2}_{B}) &=& \frac{\langle \bar{s} s \rangle \langle \bar{q} q \rangle \langle \bar{q} g_{s} \sigma \cdot G q\rangle}{216}
\int^{1}_{0} \text{d} x \frac{m_{c}^{3}( 2M_{B}^{2} - m_{c} m_{s})}{M_{B}^{6}}  e^{\frac{m_{c}^{2}}{ (x-1) M_{B}^{2}}},\\
\Pi_{A}^{\langle \bar{q} q \rangle^{2} \langle \bar{s} G s\rangle}(M^{2}_{B}) &=& \frac{\langle \bar{q} q \rangle^{2} \langle \bar{s} g_{s} \sigma \cdot G s\rangle}{648} \int^{1}_{0} \text{d} x \frac{m_{c}^{3}( 3M_{B}^{2} - m_{c} m_{s})}{M_{B}^{3}}  e^{\frac{m_{c}^{2}}{ (x-1) M_{B}^{6}}}.
\end{eqnarray}

For the sake of brevity, in this paper, we only provide the analytic results of the OPE side for the scalar case A, omitting the analytic results for other scalar cases and other quantum numbers.

On the phenomenological side, after isolating the ground state contribution of the tetraquark states, we obtain the phenomenological representation of the correlation function $\Pi_i(q^2)$, which is expressed as a dispersion integral over the hadronic spectrum,
\begin{eqnarray}
  \Pi_{i}^{\text{phen}}(q^2) = \frac{(\lambda_{X}^{2})^{2}} {(M_{X}^{i})^{2} - q^2 } +  \int_{s_0}^\infty \text{d}s \frac{\rho_{X}^{i}(s)}{s - q^2},
\end{eqnarray}
where, $M_{X}^{i}$ represents the mass of the $J^{P}=0^{+}$ tetraquark state, $\rho_{X}^{i}(s)$ denotes its spectral density that includes contributions from higher excited states and continuum states, and $s_{0}$ is the threshold for the higher excited states and continuum states. The coupling constant $\lambda_{X}$ is defined by $\langle 0|j_{X}^{i}|X\rangle = \lambda_{X}^{i}$, where $X$ is the lowest-lying tetraquark state.

Performing the Borel transformation on the phenomenological side of the correlation function, we obtain:
\begin{eqnarray}
  \Pi_{i}^{\text{phen}}(s_0, M_B^2) = (\lambda^i_{X})^{2}e^{-(M_X^i)^2/M_B^2} + \int_{s_0}^\infty \text{d}s \rho^i_X(s) e^{-s/M_B^2}.
  \label{Pi-phen}
\end{eqnarray}
According to the quark-hadron duality principle, we have the equality between equation (\ref{Pi-OPE-MBMB}) and equation (\ref{Pi-phen}), thus:
\begin{eqnarray}
  &&(\lambda^{i}_{X})^{2}e^{-(M_X^i)^2/M_B^2} + \int_{s_0}^\infty \text{d}s \rho^i_X(s) e^{-s/M_B^2}  \nonumber \\
  &=& \int_{(m_s + m_c)^2}^\infty \text{d}s \rho^{\text{OPE}}_i(s)e^{-s/M_B^2} + \Pi_{i}^{\langle G^{3} \rangle}(M^{2}_{B})
  +\Pi_{i}^{\langle G^{2} \rangle \langle \bar{s} s \rangle}(M^{2}_{B}) + \Pi_{i}^{\langle \bar{s} s \rangle \langle \bar{q} G q \rangle }(M^{2}_{B})+ \cdots .\label{quark-hadron-duality}
\end{eqnarray}

Under the quark-hadron duality approximation, for the equation (\ref{quark-hadron-duality}), the integral on the left side is approximately equal to the integral of the first term on the right side in the interval $s_0 < s <\infty$. Therefore, both sides can simultaneously remove this integral part, ultimately resulting in the relationship between the ground state and the OPE calculation results:
\begin{eqnarray}
  (\lambda^{i}_{X})^{2}e^{-(M_X^i)^2/M_B^2} &=&\int_{(m_s + m_c)^2}^{s_0} \text{d}s \rho^{\text{OPE}}_i(s)e^{-s/M_B^2} + \Pi_{i}^{\langle G^{3} \rangle}(M_B^2) \nonumber \\
  & &+\Pi_{i}^{\langle G^{2} \rangle \langle \bar{s} s \rangle}(M^{2}_{B}) + \Pi_{i}^{\langle \bar{s} s \rangle \langle \bar{q} G q \rangle }(M^{2}_{B}) + \cdots .
  \label{Pi-OPE=Pi-phen}
\end{eqnarray}

For convenience, we define the zeroth moment in the QCD sum rules, which equals the right side of the aforementioned equation, i.e.,
\begin{eqnarray}
  L_0^i(s_0, M_B^2) &=& \int_{m_{s}^{2}}^\infty \text{d}s \, \rho^{\text{OPE}}(s) e^{-s/M_B^2} \nonumber\\
  &+& \Pi_{i}^{\langle G^{3} \rangle}(M_B^2) + \Pi_{i}^{\langle G^{2} \rangle \langle \bar{s} s \rangle}(M_B^2) + \Pi_{i}^{\langle \bar{s} s \rangle \langle \bar{q} G q \rangle }(M_B^2) +\cdots \,.
\end{eqnarray}

Taking the derivative of the zeroth moment $L_0^i(s_0, M_B^2)$ with respect to $1/M_B^2$, we obtain the first moment $L_1^i(s_0, M_B^2 )$:
\begin{eqnarray}
  L_1^{i}(s_0, M_B^2) \!\!=\!\! \frac{\partial}{\partial (M_B^2)^{-1}} L_0^i(s_0, M_B^2).
\end{eqnarray}

From the left side expression of equation (\ref{Pi-OPE=Pi-phen}), it is evident that $-(M^i_X)^2 = L_1(s_0, M_B^2)/L_0(s_0, M_B^2)$, thus the mass of the studied hadron is:
\begin{eqnarray}
  M_{X}^{i}(s_{0},M_{B}^{2}) = \sqrt{-\frac{L_1^{\text{i}}(s_0, M_B^2)}{L_0^i(s_0, M_B^2)}}. \label{mass-equation}
\end{eqnarray}

In this paper, we will determine the mass of the studied tetraquark states by analyzing equation (\ref{mass-equation}).

\section{Numerical analysis}\label{sec:numerical}
In this section, we take the current $j_{A}^{0^+}$ as an example and perform the numerical analysis. In our analysis, we utilize the quark masses and various QCD parameters as provided in Table~\ref{QCDParam} \cite{Reinders:1984sr, Shifman:1978bx, Shifman:1978by, Narison:1989aq, Colangelo:2000dp, ParticleDataGroup:2022pth}, where $\overline{m_c}$ represents the running mass of the charm quark in the $\overline{\text{MS}}$ scheme.
\begin{table}[h]
\begin{center}
\setlength{\tabcolsep}{1.25pc}
\caption{~~~~The QCD input parameters required for the numerical calculations in this paper~\cite{Reinders:1984sr, Shifman:1978bx, Shifman:1978by, Narison:1989aq, Colangelo:2000dp, ParticleDataGroup:2022pth}.}
\begin{tabular}{ll}
&\\
\hline
Parameter Name & Value\\
\hline
$m_u,m_d$ & $0$ \\
$m_c(m_c)=\overline{m_c}$ & $(1.27\pm0.02)~\text{GeV}$ \\
$m_s$ & $(0.13 \pm 0.03)~\text{GeV}$ \\
$\langle q \bar{q}\rangle$ & $-(0.24 \pm 0.01)^3~\text{GeV}^3$\\
$\langle g_s^2 G^2\rangle$ & $0.88~\text{GeV}^4$\\
$\langle g_s^3 G^3\rangle$ & $0.045~\text{GeV}^6$\\
$\langle s \bar{s}\rangle / \langle q \bar{q}\rangle$ & $(0.8 \pm 0.1)$\\
$m_0^2 \equiv \langle qG \bar{q}\rangle / \langle q \bar{q}\rangle$ & $(0.8\pm0.2)~\text{GeV}^2$\\
\hline
\end{tabular}
\label{QCDParam}
\end{center}
\end{table}

As mentioned earlier (\ref{mass-equation}), the extracted hadron mass $M_{X}$ is a function of the threshold parameter $s_{0}$ and the Borel parameter $M_{B}^2$, which are two important parameters in QCD sum rule analysis. If the final result $M_{X}$ does not depend on these two free parameters, then the QCD sum rule method will have perfect predictive power. However, in reality, reliable mass predictions can be obtained when these parameters exhibit weak dependencies within a reasonable working region. Determining the Borel window (a reasonable working region for $M_{B}^2$) primarily involves two criteria: the requirement of OPE convergence provides the lower limit of the parameters $(M_B^2)_{\text{min}}$, while the dominance of the pole contribution determines the upper limit $(M_B^2)_{\text{max}}$. Additionally, we need to analyze the variation of the hadron mass $M_{X}$ with the threshold parameter $s_0$. To minimize the dependence of the extracted hadron mass $M_{X}$ on the Borel mass $M_{B}^2$, an optimal threshold parameter $s_{0}$ needs to be chosen, where the $M_X-M_B^2$ curve exhibits a good plateau.

In the analytic calculations of this paper, we include non-perturbative terms, i.e., condensates, up to the dimension-11 terms. Through numerical analysis, we find that the contributions of the two-quark condensate $\langle q\bar{q}\rangle$ and the four-quark condensate $\langle q\bar{q}\rangle^{2}$ are the dominant non-perturbative contributions, while the contributions of other condensates are much smaller. For the additional parameters $M_{B}^{2}$ and $s_{0}$ introduced by QCD sum rules, we will constrain these parameters according to the two fundamental criteria introduced earlier. For instance, in the references \cite{Shifman:1978bx,Shifman:1978by,Reinders:1984sr,Colangelo:2000dp}, the authors elaborate on these two criteria. We outline these criteria as follows in this paper: The first criterion is OPE convergence, which requires comparing the ratio of each condensate contribution to the total contribution and selecting a reliable $M_{B}^{2}$ region to maintain their convergence. In practical applications, the degree of OPE convergence can be expressed by the following formula:
\begin{eqnarray}
  R_{i}^{\text{OPE}} (s_0, M_B^2)= \frac{L_0^{\text{cond}}(s_0, M_B^2)}{L_0(s_0, M_B^2)}\,.
\end{eqnarray}
The second criterion is the pole contribution (also known as the ground state contribution) \cite{Reinders:1984sr,Colangelo:2000dp}, which compares the ratio of the ground state contribution to the total hadron spectrum contribution. Generally, this ratio should be greater than $50\%$, but sometimes, depending on the specific problem, this percentage can be slightly relaxed, but it still needs to be greater than $40\%$, and can be expressed as:
\begin{eqnarray}
  R_{i}^{\text{PC}} (s_0, M_B^2)= \frac{L_0(s_0, M_B^2)}{L_0(\infty, M_B^2)} \; . \label{RatioPC}
\end{eqnarray}
By constraining these parameters with these two criteria, we can reasonably eliminate contributions from not only higher-order condensates but also higher excited states and continuum states.

At the same time, in order to find a suitable value for $\sqrt{s_{0}}$, we conducted a scan. Since the threshold parameter $s_{0}$ is associated with the mass of the ground state through $\sqrt{s_{0}} \sim (M_{X} + \delta) $ GeV, where the range of $\delta $ is between 0.4 and 0.8 GeV, various values of $ \sqrt{s_{0}} $ that satisfy this constraint should be considered in the numerical analysis. Among these values, we aim to identify one that offers an optimal window for the Borel parameter $ M_{B}^{2} $, or in other words, where the mass of the tetraquark state $ M_{X} $ is as independent of the Borel parameter $ M_{B}^{2} $ as possible. Ultimately, the value of $ \sqrt{s_{0}} $ corresponding to the optimal mass curve becomes the central value of $ \sqrt{s_{0}} $. In practical applications, that is, in the calculations of QCD sum rules, a variation of $ \sqrt{s_{0}} $ by 0.2 GeV is acceptable, thus setting the upper and lower limits for $ \sqrt{s_{0}} $. The fluctuation of the parameter $ \sqrt{s_{0}} $ then corresponds to its range of error.

\begin{figure}
\begin{center}
\includegraphics[width=7.5cm]{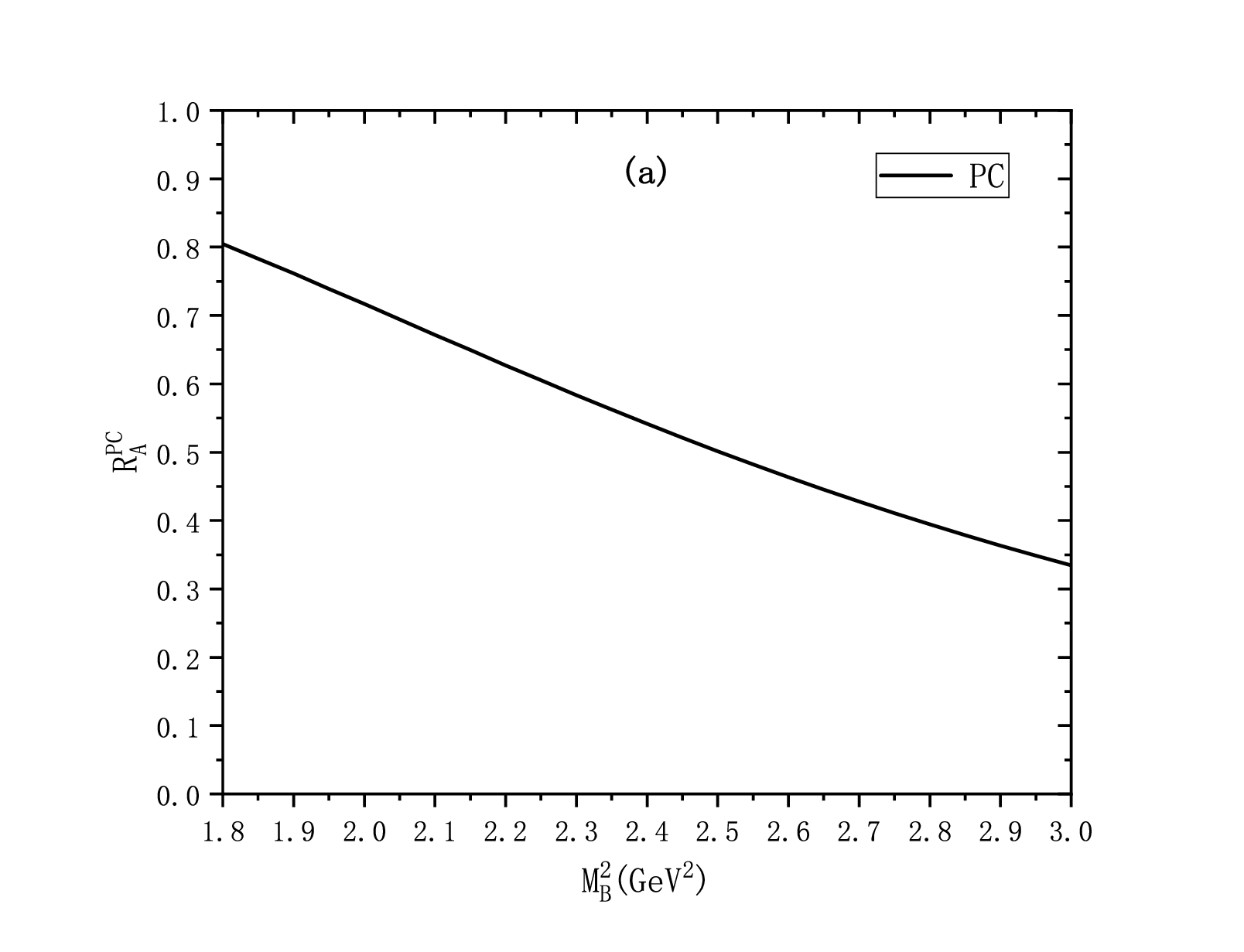}
\includegraphics[width=7.5cm]{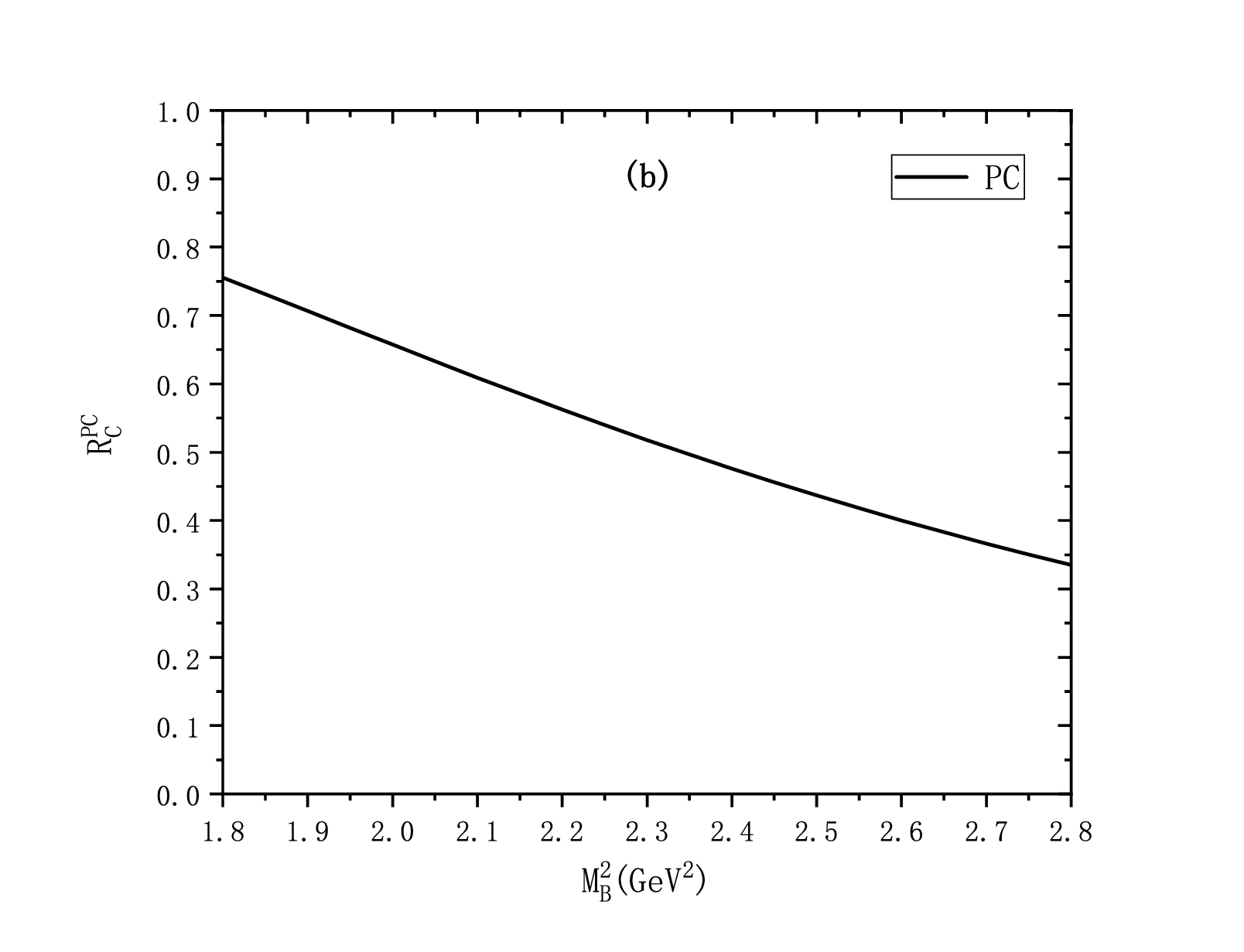}
\caption{~~~For the case of $J^{P} = 0^{+}$, graphs showing the variation of the pole contribution ratio with the Borel parameter $M_B^2$. Figure (a) shows the $R_{A}^{\text{PC}}-M_B^2$ curve; figure (b) shows the $R_{C}^{\text{PC}}-M_B^2$ curve, where $s_{0}$ corresponds to the optimal threshold parameter obtained after scanning, with $s_{0} = 3.6^{2} $ GeV$^{2}$, and the x-axis represents the Borel parameter $M^{2}_{B}$. }
\label{PC0+}
\end{center}
\end{figure}

\begin{figure}
\begin{center}
\includegraphics[width=7.5cm]{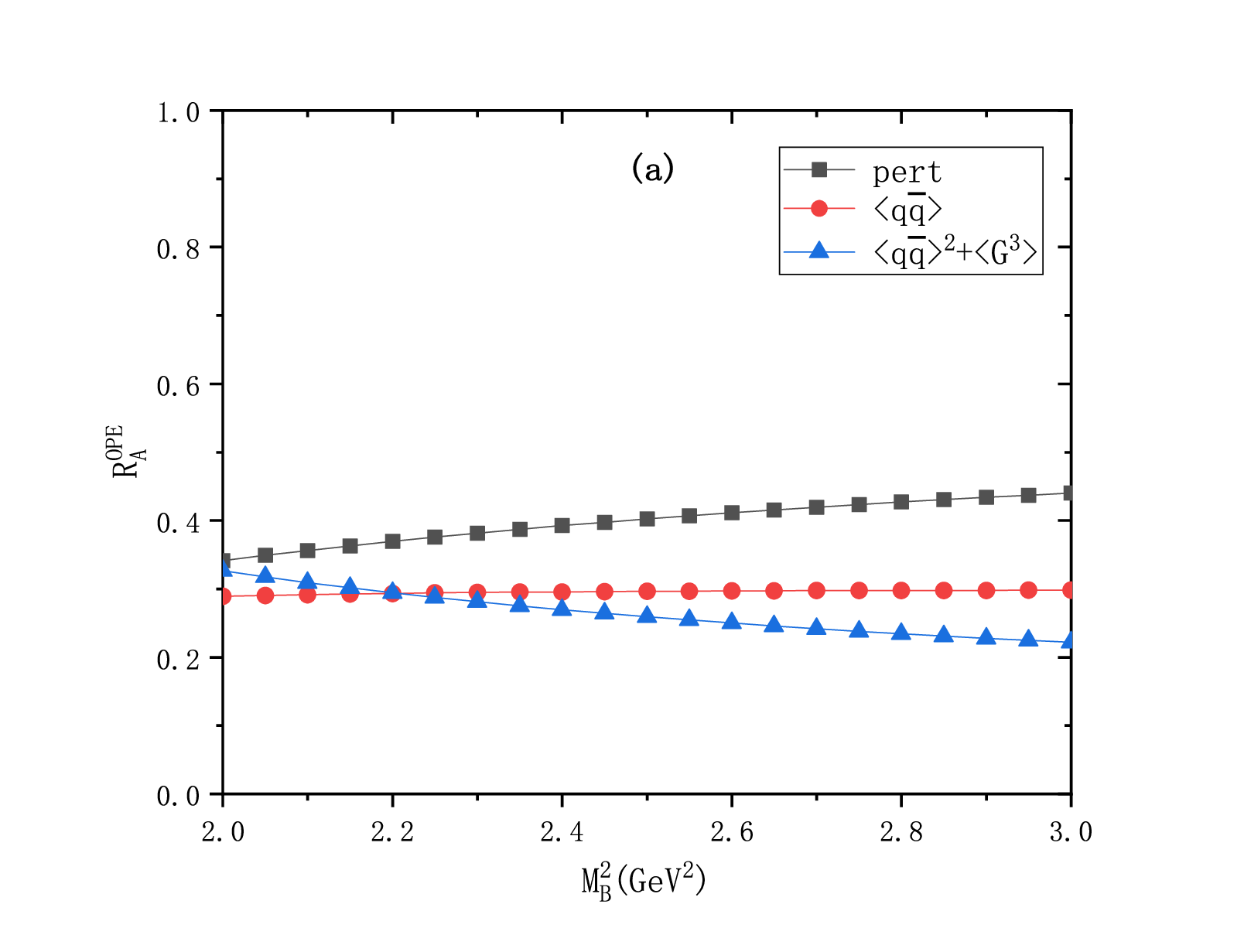}
\includegraphics[width=7.5cm]{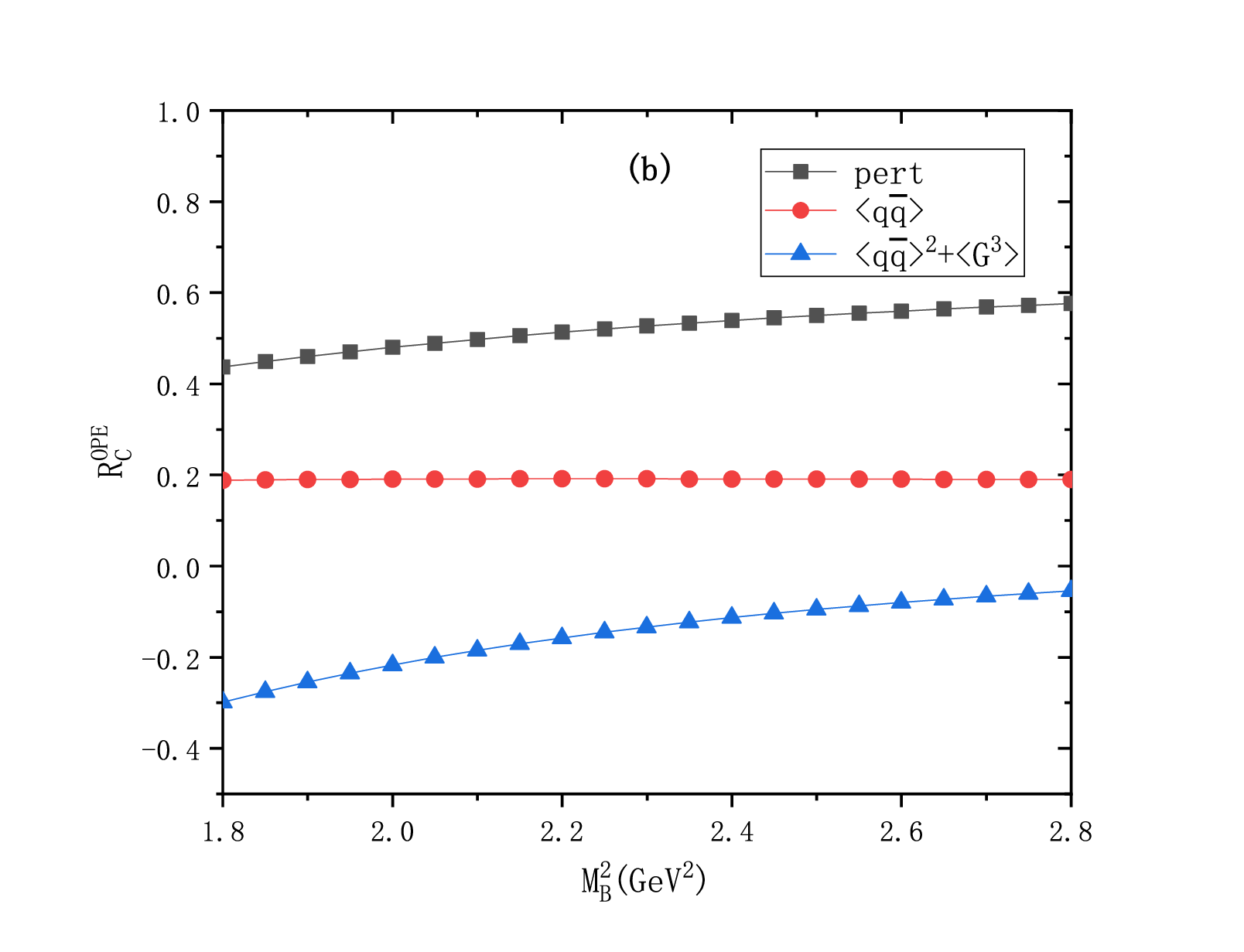}
\caption{~~~For $J^{P} = 0^{+}$, the OPE convergence ratio $R_{A/C}^{\text{OPE}}$ as a function of the parameter $M_B^2$, where $s_{0}$ corresponds to the optimal threshold parameter obtained after scanning, $s_{0} = 3.6^{2} $GeV$^{2}$. Figure (a) corresponds to current (\ref{current-A}), and figure (b) corresponds to current (\ref{current-C}). In the graphs, the solid line with black squares represents the ratio of perturbations, the red solid line with red circles represents the ratio of two-quark condensates $\langle q\bar{q}\rangle$, and the blue solid line with blue triangles represents the ratio of four-quark condensates $\langle q\bar{q}\rangle^{2}$. These three ratios are shown because, apart from perturbations, only the two-quark and four-quark condensates make significant contributions, with other contributions being very small, hence not marked on the graph with their contribution ratios. }
\label{ROPE0+}
\end{center}
\end{figure}

\begin{figure}[htp]
\begin{center}
\includegraphics[width=7.5cm]{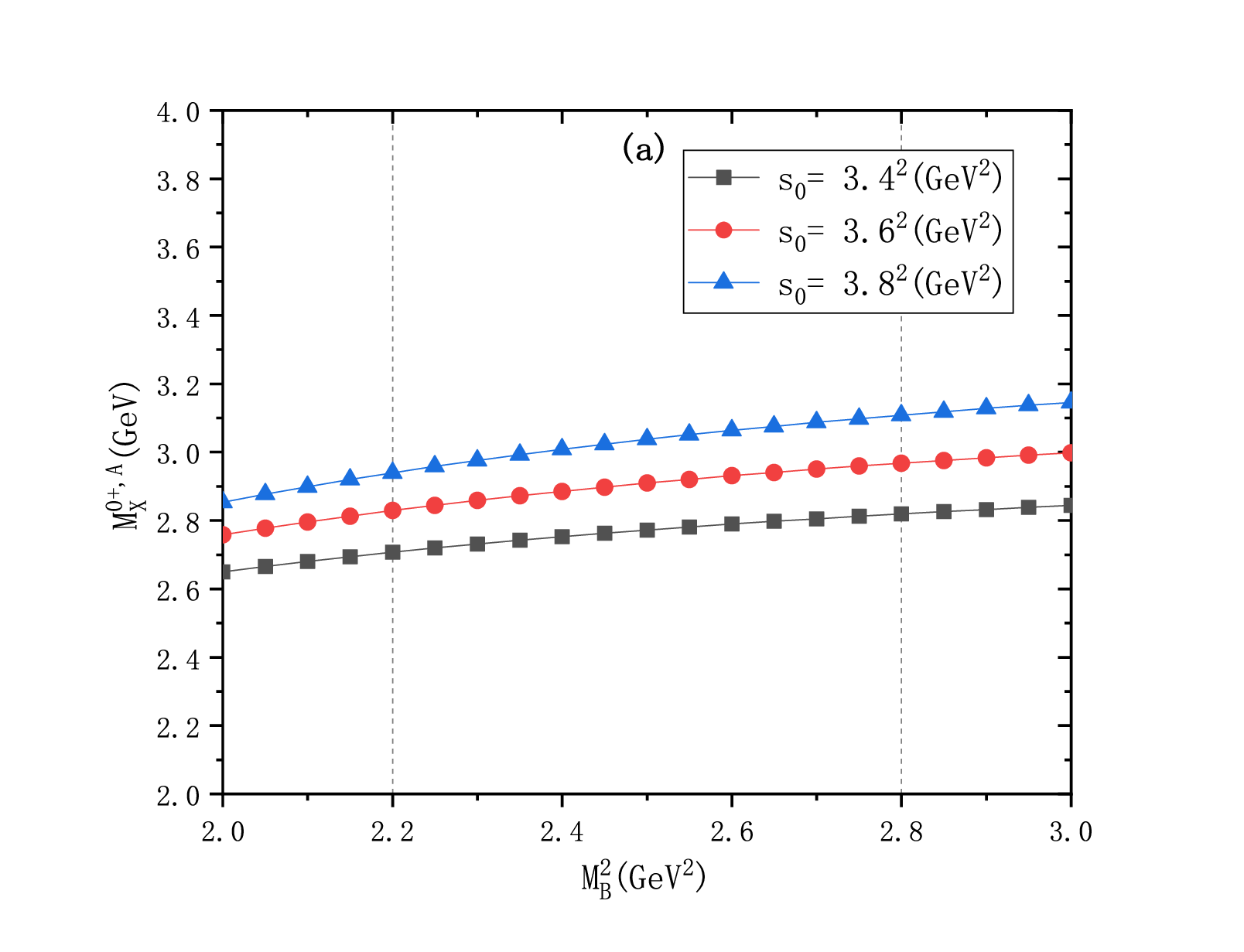}
\includegraphics[width=7.5cm]{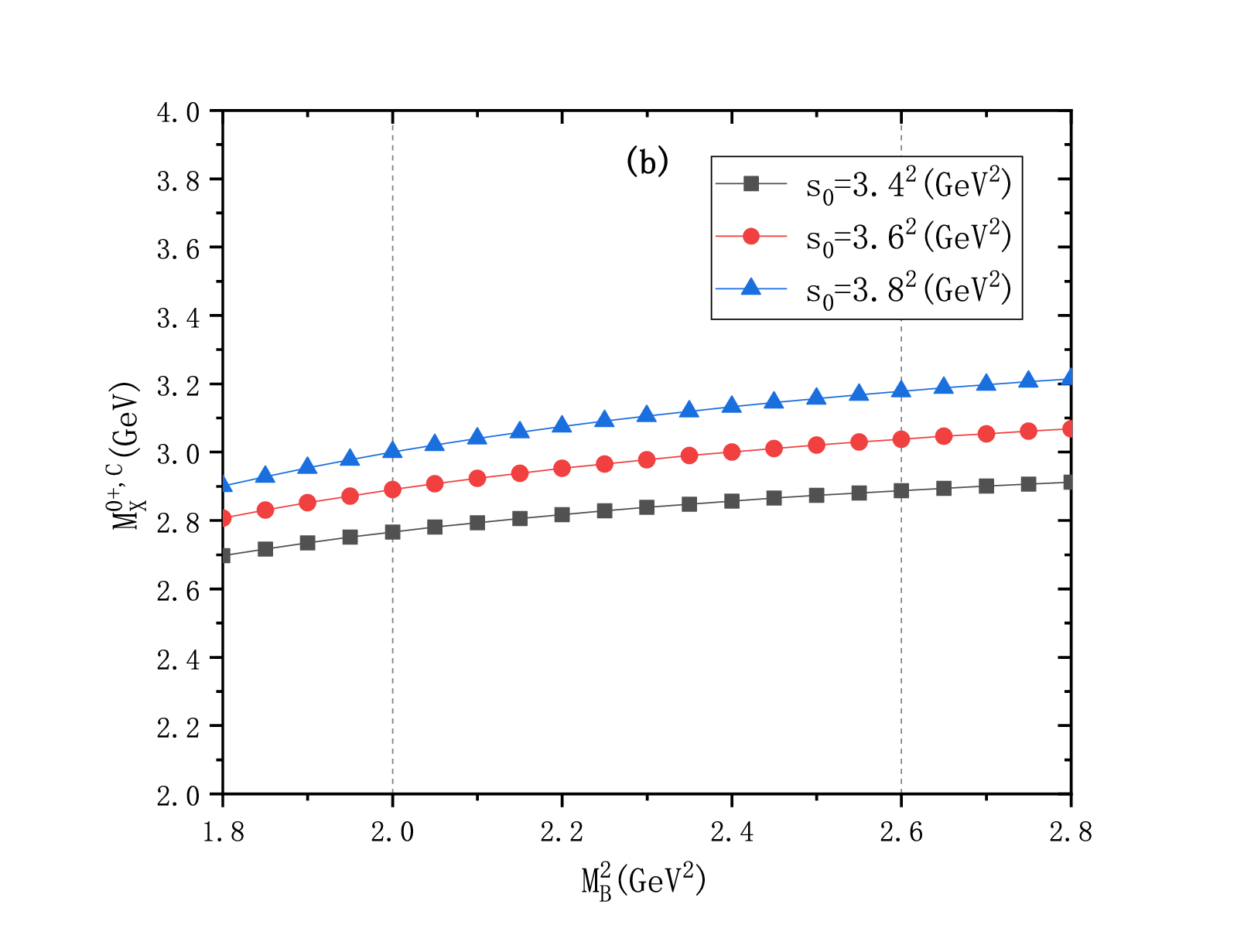}
\caption{~~~For $J^{P} = 0^{+}$ under different threshold parameters $s_0$, the mass $M^{A/C}_{X}$ as a function of $M^{2}_{B}$, with figure (a) corresponding to current (\ref{current-A}), and figure (b) corresponding to current (\ref{current-C}). In each graph, the two vertical dashed lines represent the upper and lower bounds of the Borel window, with the central value of $s_{0}$ as the input.} \label{mass0+}
\end{center}
\end{figure}

In Fig.\ref{PC0+}, we plot the pole contribution $R^{\text{PC}}$ curves for currents with quantum numbers $J^P=0^+$, specifically for currents (\ref{current-A}) and (\ref{current-C}). The upper limit constraints on the Borel parameter $M_B^{2}$ obtained were $M_B^{2}=2.8~\text{GeV}^{2}$ and $2.6~\text{GeV}^{2}$, respectively, where the threshold parameter $s_0$ used was the optimal value of $3.6^{2}~\text{GeV}^2$. The corresponding OPE convergence curves for the two currents are shown in Fig.\ref{ROPE0+}, where the lower limits for $M^{2}_{B}$ were found to be $M^{2}_{B}=2.2~\text{GeV}^{2}$ and $2.0~\text{GeV}^2$, again with the threshold parameter $s_0$ set to the optimal value of $3.6^{2}~\text{GeV}^2$. It's important to note that the constraints on the upper and lower limits of $M^{2}_{B}$ also depend on the threshold parameter $s_{0}$. That is, for different values of $s_{0}$, $M^{2}_{B}$ will have different upper and lower limits. To determine the optimal $s_{0}$ value, we performed an analysis similar to that found in references \cite{Shifman:1978bx, Shifman:1978by, Reinders:1984sr, Colangelo:2000dp}. For different values of $s_{0}$, the masses $M_{X}^{A}$ and $M_{X}^{C}$ as functions of the Borel parameter $M^{2}_{B}$ are shown in Fig.\ref{mass0+}. In analyzing the cases of currents (\ref{current-B}) and (\ref{current-D}), we find that no reasonable Borel parameter $M_B^2$ window could be obtained regardless of the value of $s_0$. Therefore, we conclude that currents (\ref{current-A}) and (\ref{current-C}) correspond to reasonable tetraquark states, with specific numerical results summarized in Table~\ref{tab0+}; currents (\ref{current-B}) and (\ref{current-D}), due to the lack of a reasonable parameter window, do not correspond to reasonable tetraquark states.


\begin{table}[h]
\centering
\caption{~~~The numerical results for the parameter $M_B^2$, threshold $\sqrt{s_0}$, PC ratio, and mass associated with currents $j_A^{0^+}$ and $j_C^{0^+}$.}
\begin{tabular}{ccccc}\hline
  $j^{J^{P}}_{X}$ & $M_B^2 \, (\rm{GeV^{2}})$  & $\sqrt{s_{0}} \, (\rm{GeV})$ & PC & Mass \, $(\rm{GeV})$  \\ \hline
  $j^{0^{+}}_{A}$ & ${2.20\!-\!2.80}$ & {$3.6 \pm 0.2$} & ${(65\!-\!40)\%}$  & ${2.91_{-0.20}^{+0.20}}$ \\ \hline
  $j^{0^{+}}_{C}$ & ${2.00\!-\!2.60}$ & {$3.6 \pm 0.2$} & ${(66\!-\!40)\%}$  & ${2.98_{-0.21}^{+0.20}}$ \\ \hline
\end{tabular}
\label{tab0+}
\end{table}

Ultimately, the masses and uncertainties for the tetraquark states with color octet-octet structure corresponding to currents $j_{A}^{0^+}$ and $j_{C}^{0+}$ are as follows:
\begin{eqnarray}
  M_X^A &=& 2.91_{-0.20}^{+0.20}~\text{GeV}, \\
  M_X^C &=& 2.98_{-0.21}^{+0.20}~\text{GeV},
\end{eqnarray}
where, the central values of the masses $M_{X}^A$ and $M_X^C$ correspond to the most stable positions on the mass curve, with errors arising from various condensates, quark masses, the threshold parameter $\sqrt{s_{0}}$, and the Borel parameter $M_{B}^{2}$.

Comparing with experimental result, our findings are consistent within the error range with the experimentally observed $T_{c\bar{s}0}^a(2900)^{++}$. Therefore, our calculations support classifying $T_{c\bar{s}0}^a(2900)^{++}$ as a tetraquark state with the color octet-octet configuration.

\begin{figure}[!ht]
\begin{center}
\includegraphics[width=7.5cm]{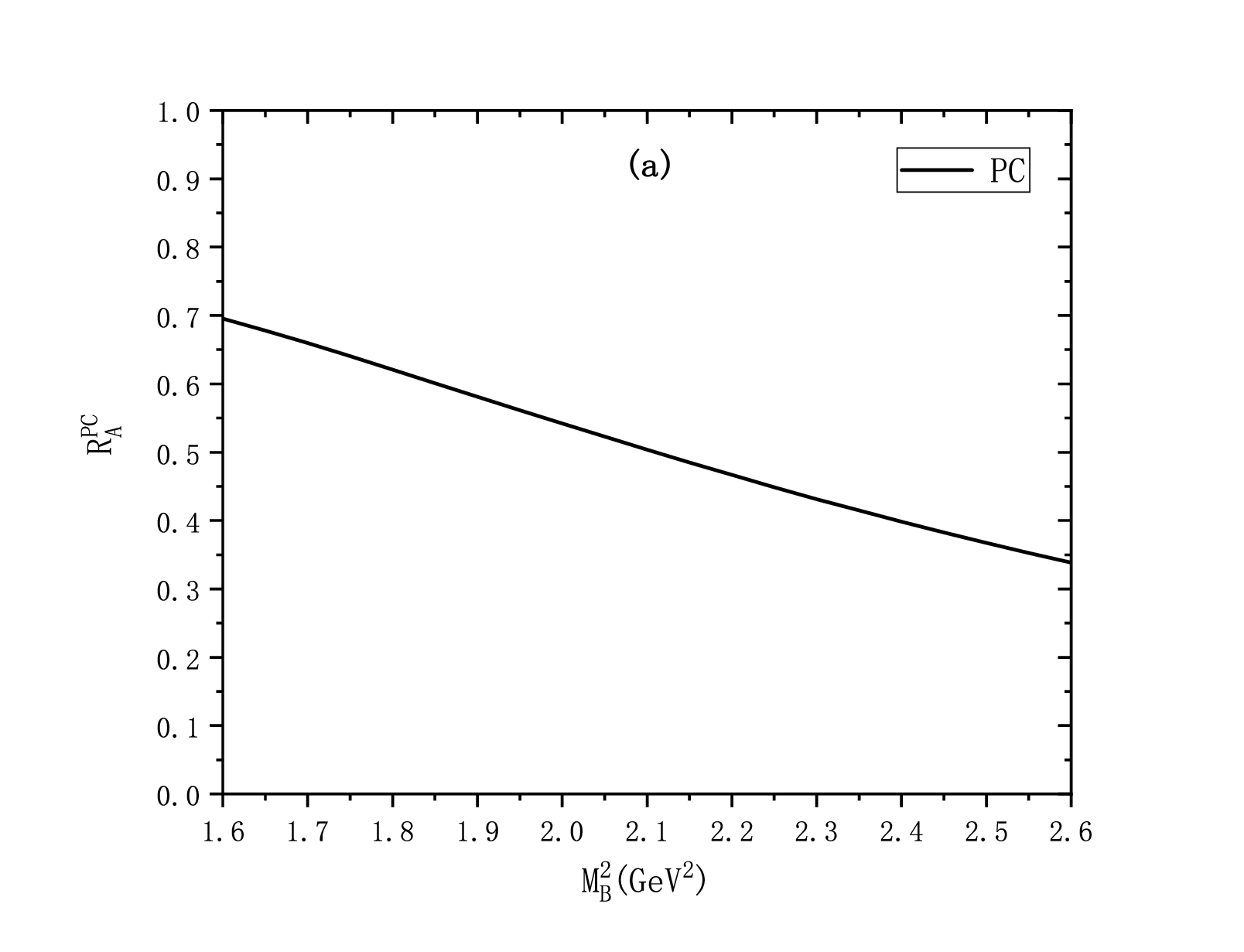}
\includegraphics[width=7.5cm]{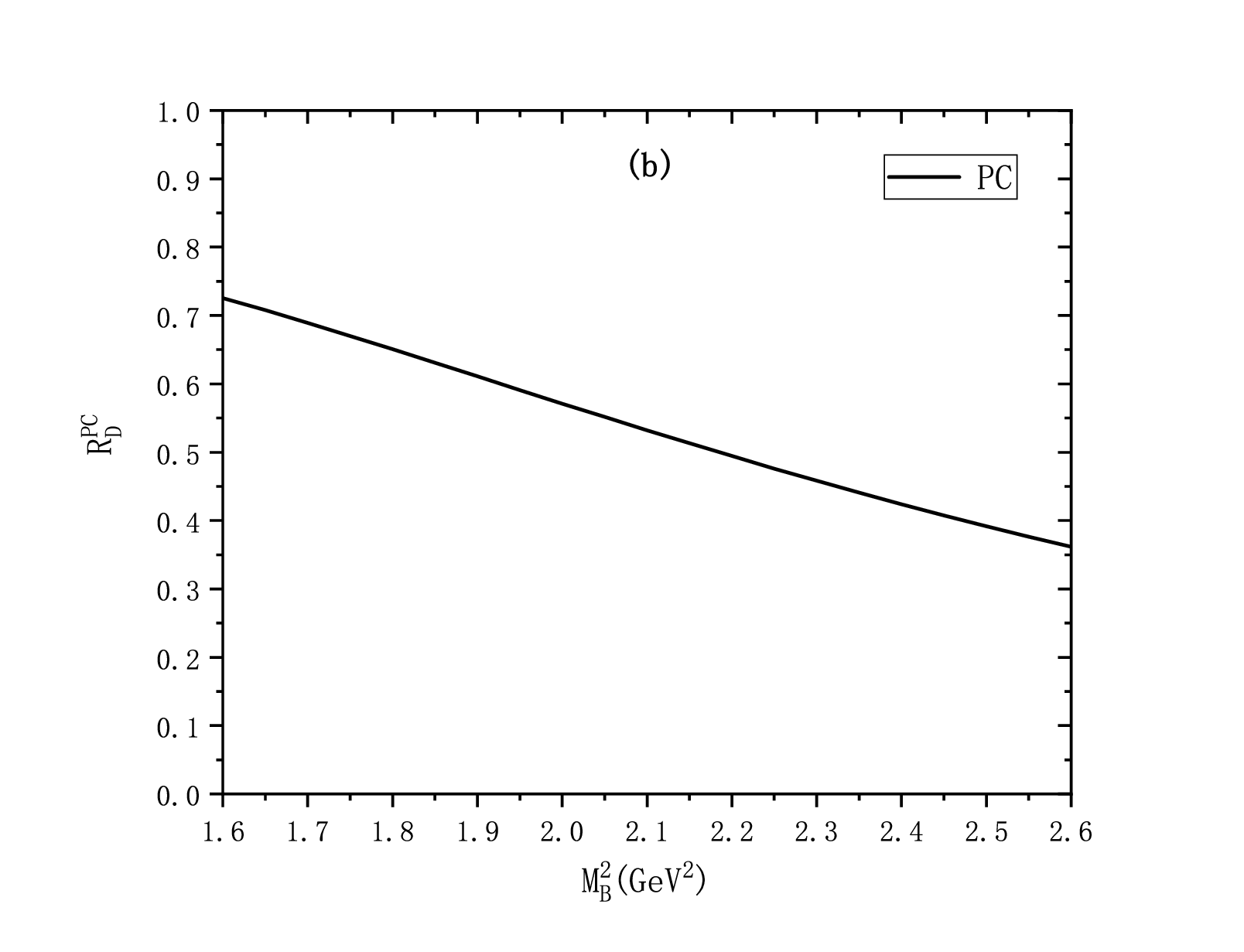}
\caption{~~~The same caption as in Fig.\ref{PC0+}, but for case $1^-$.}
\label{PC1-}
\end{center}
\end{figure}

\begin{figure}[!ht]
\begin{center}
\includegraphics[width=7.5cm]{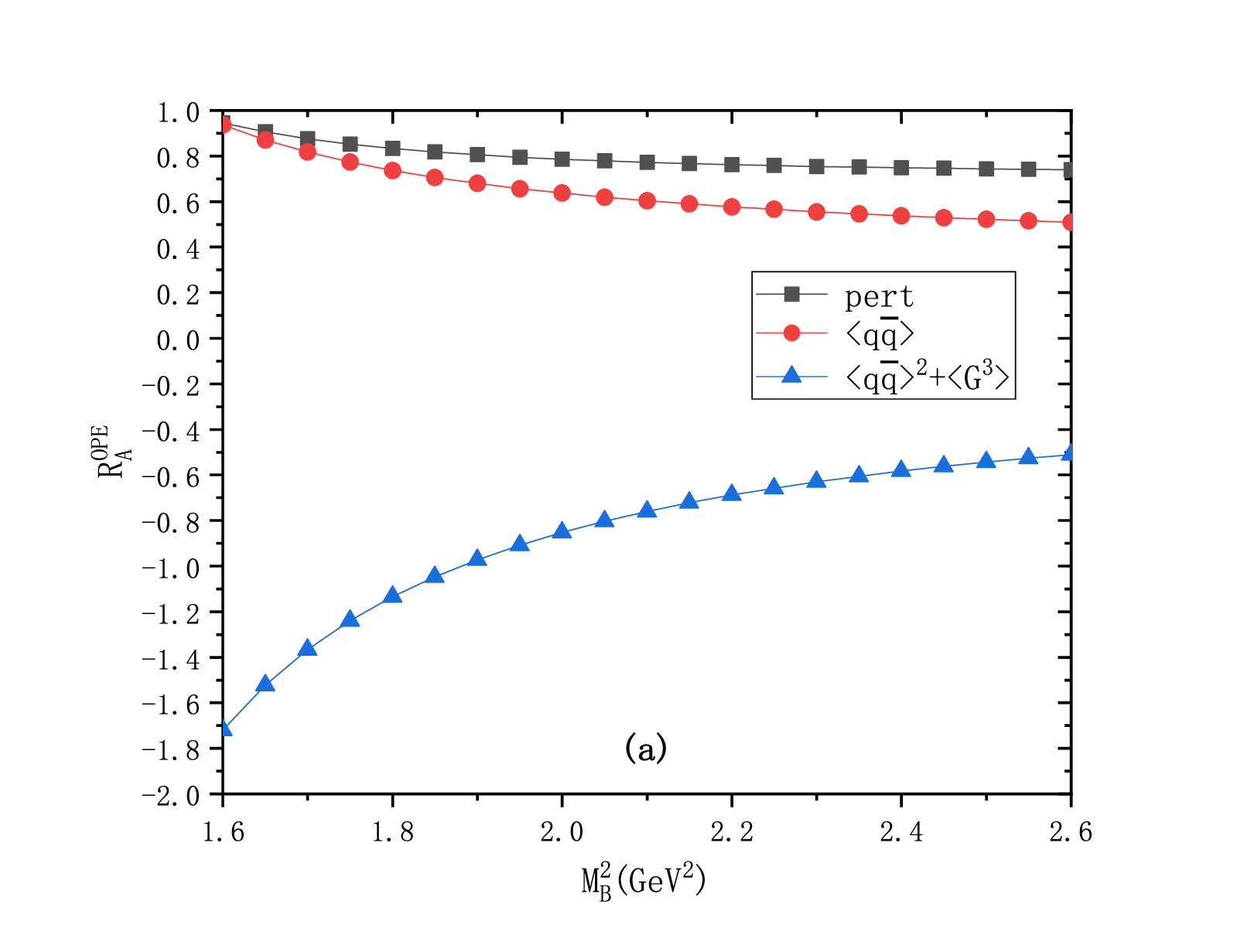}
\includegraphics[width=7.5cm]{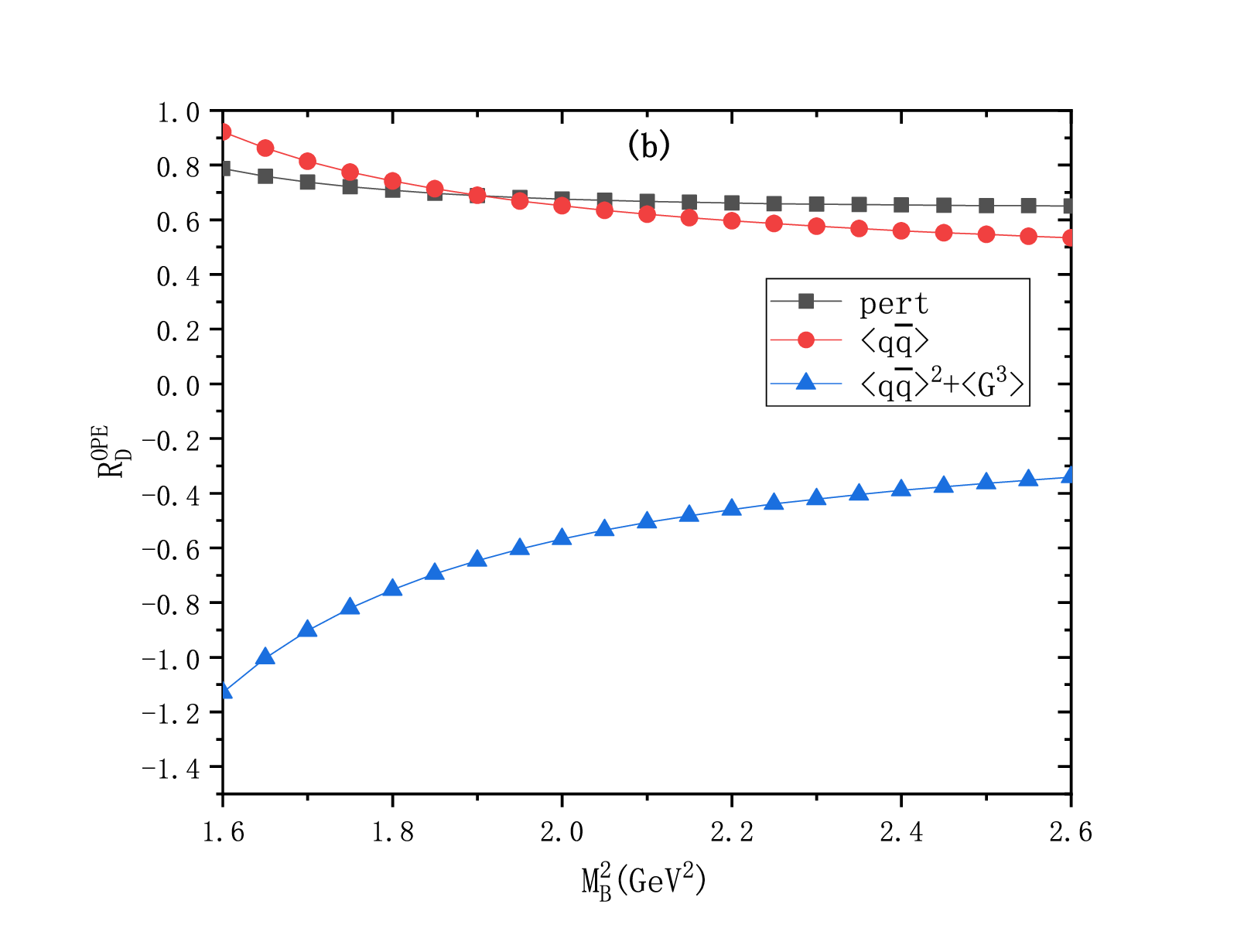}
\caption{~~~The same caption as in Fig.\ref{ROPE0+}, but for case $1^-$. }
\label{ROPE1-}
\end{center}
\end{figure}

\begin{figure}[!ht]
\begin{center}
\includegraphics[width=7.5cm]{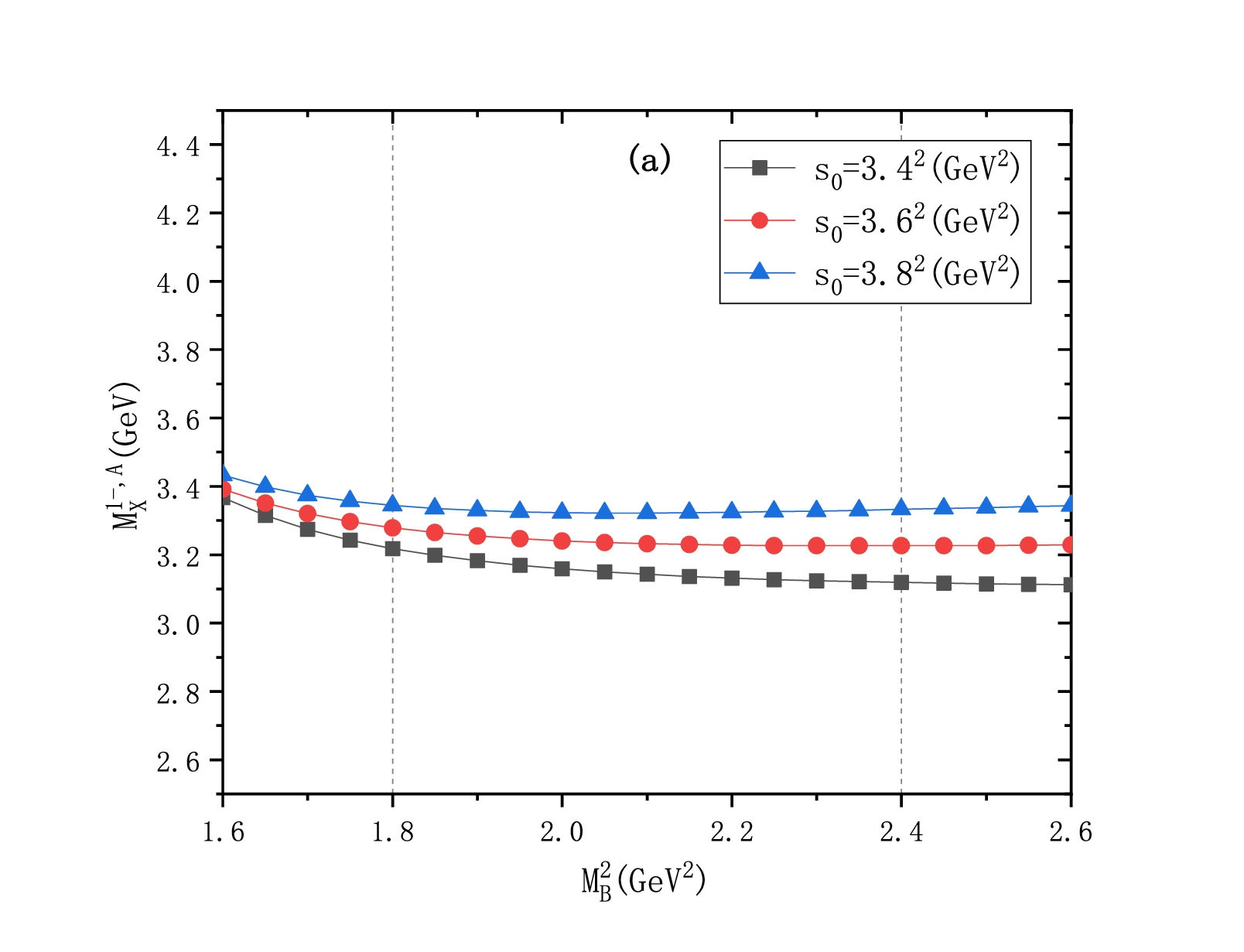}
\includegraphics[width=7.5cm]{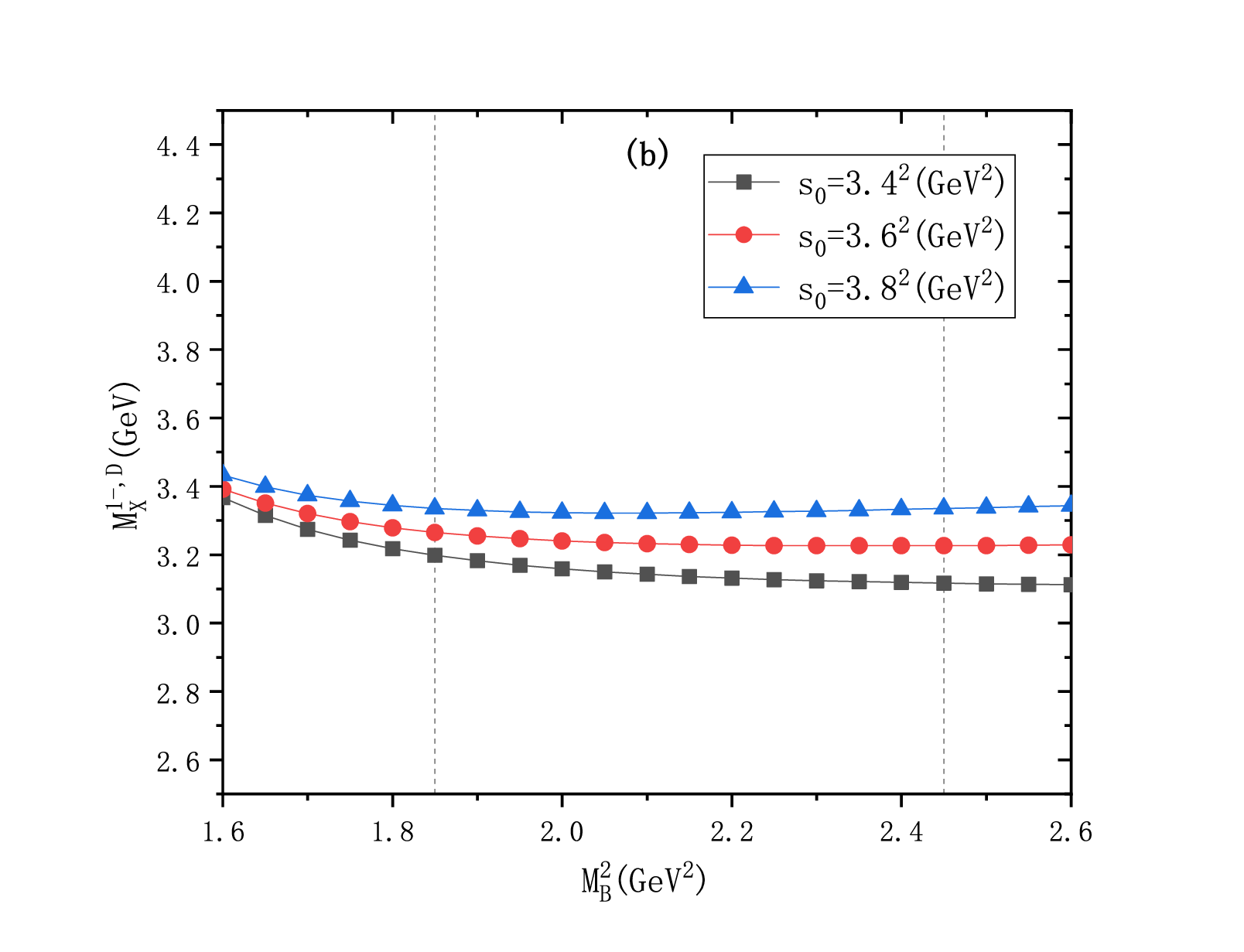}
\caption{~~~The same caption as in Fig.\ref{mass0+}, but for case $1^-$.}
\label{mass1-}
\end{center}
\end{figure}

In the following, we present the PC ratio plots, OPE ratio plots, and mass curve plots for the quantum numbers $0^-$, $1^+$, and $1^-$. The following Figs.\ref{PC0-}-\ref{mass0-},  \ref{PC1+}-\ref{mass1+}, and \ref{PC1-}-\ref{mass1-} display the numerical results for cases $0^-$, $1^+$, and $1^-$, respectively.

\begin{figure}[!ht]
\begin{center}
\includegraphics[width=7.5cm]{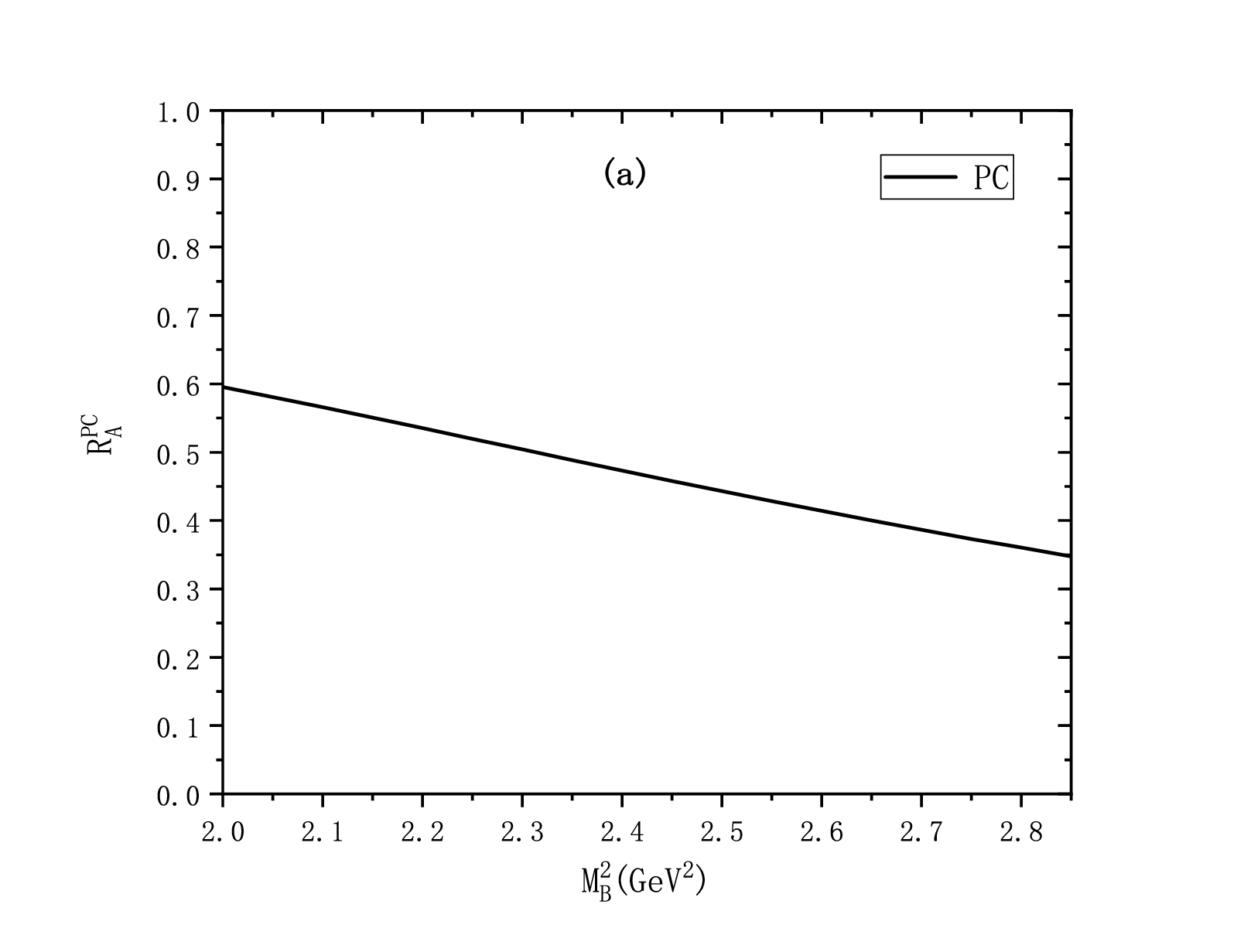}
\includegraphics[width=7.5cm]{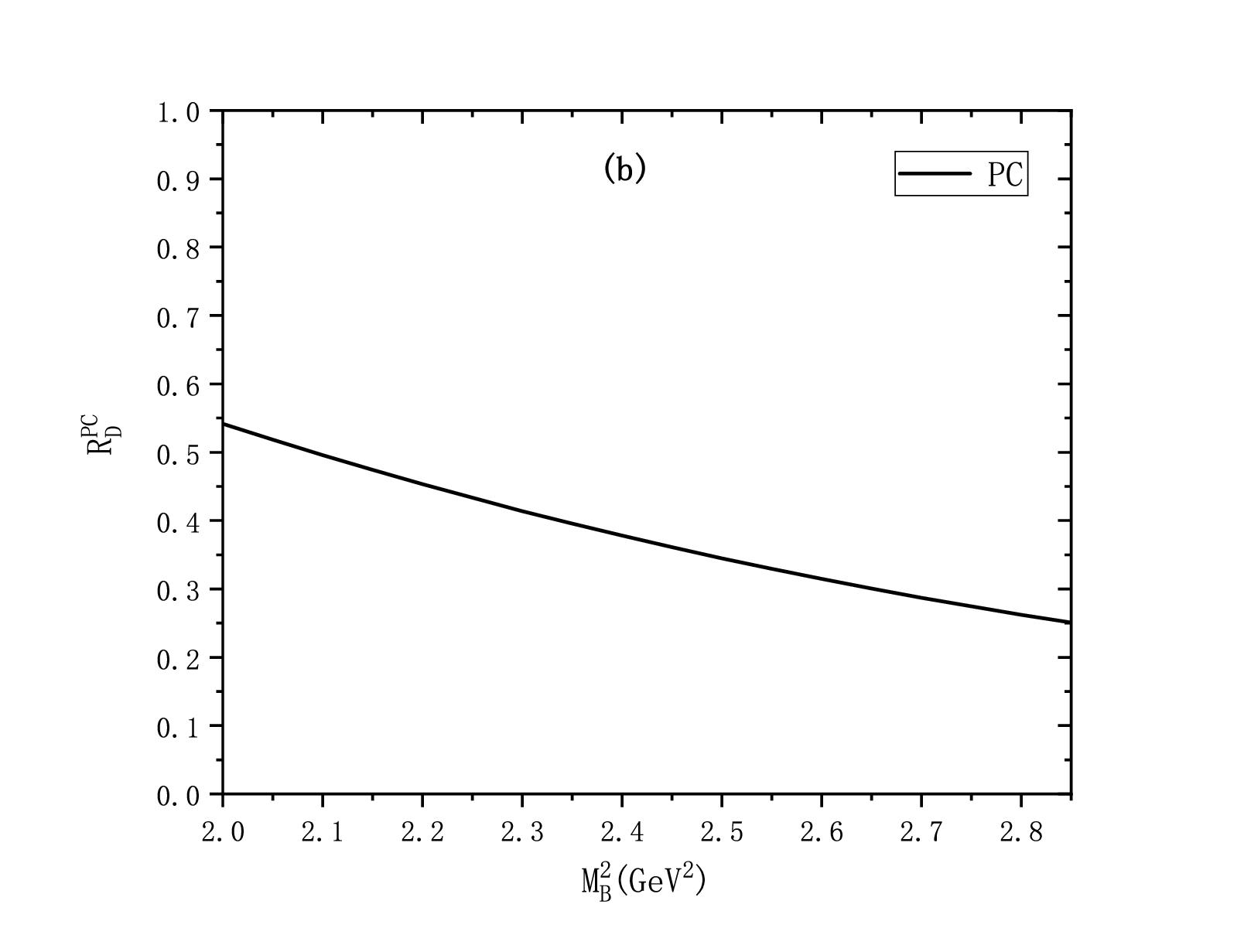}
\caption{~~~The same caption as in Fig.\ref{PC0+}, but for case $0^-$. }
\label{PC0-}
\end{center}
\end{figure}

\begin{figure}[!ht]
\begin{center}
\includegraphics[width=7.5cm]{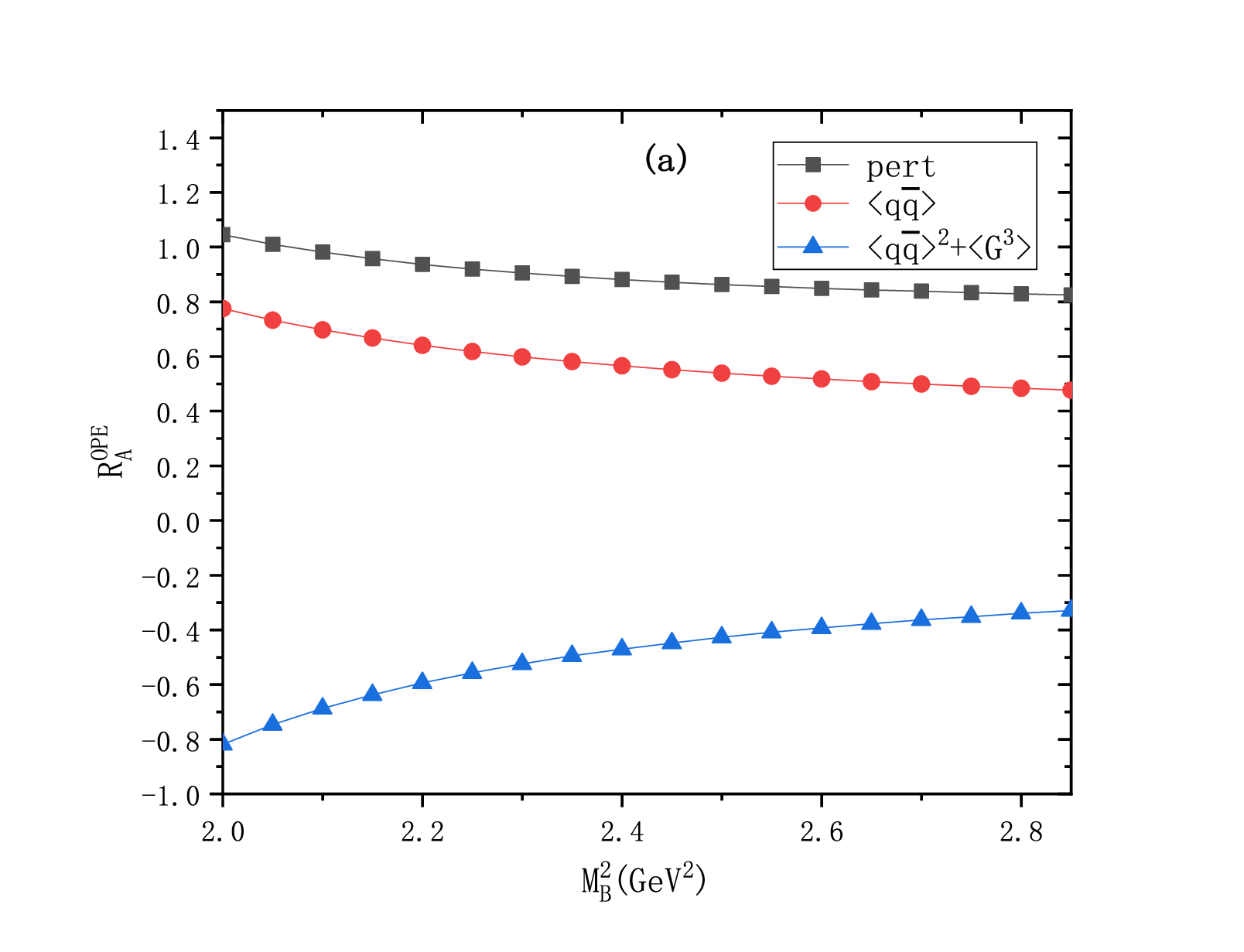}
\includegraphics[width=7.5cm]{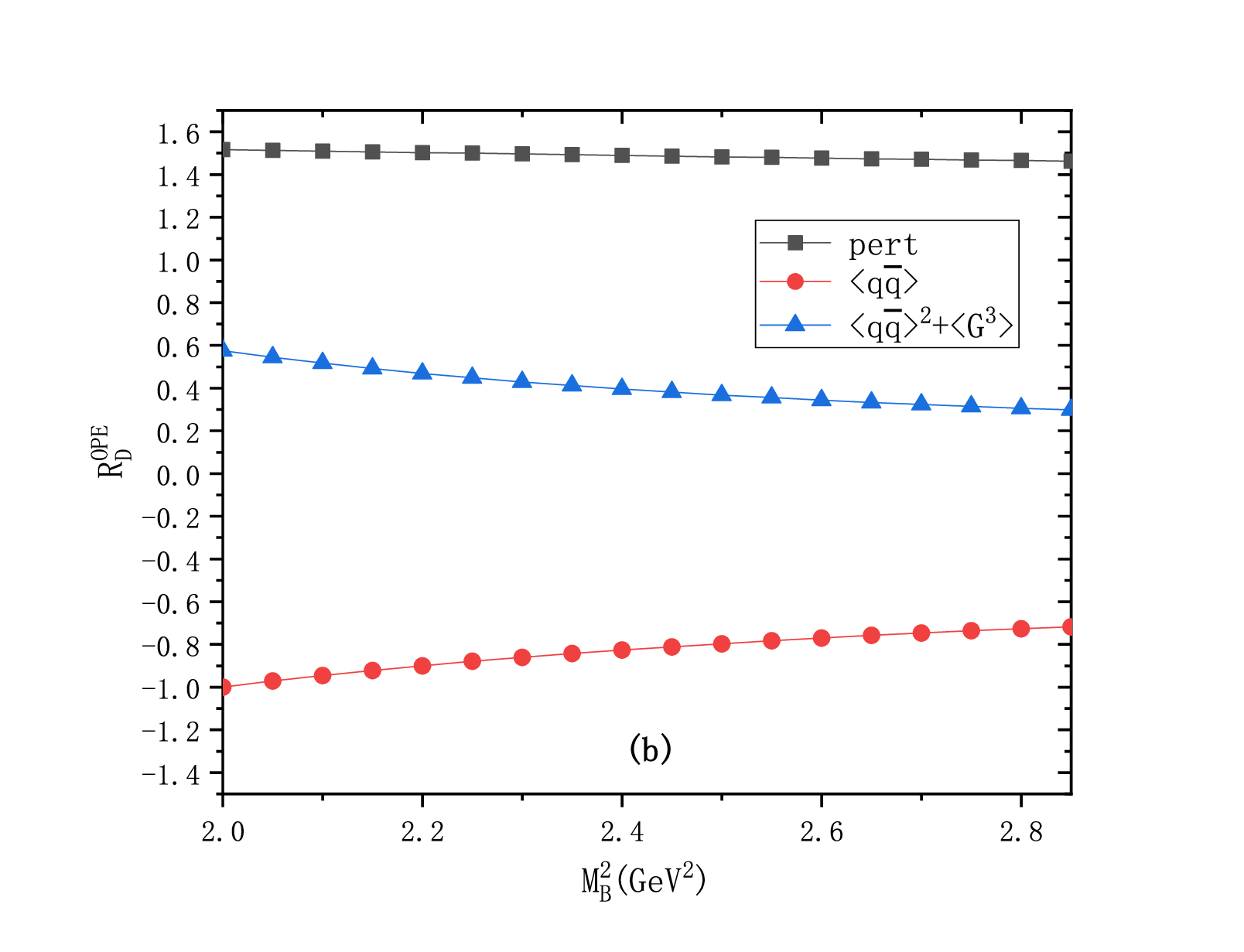}
\caption{~~~The same caption as in Fig.\ref{ROPE0+}, but for case $0^-$. }
\label{ROPE0-}
\end{center}
\end{figure}

\begin{figure}[!ht]
\begin{center}
\includegraphics[width=7.5cm]{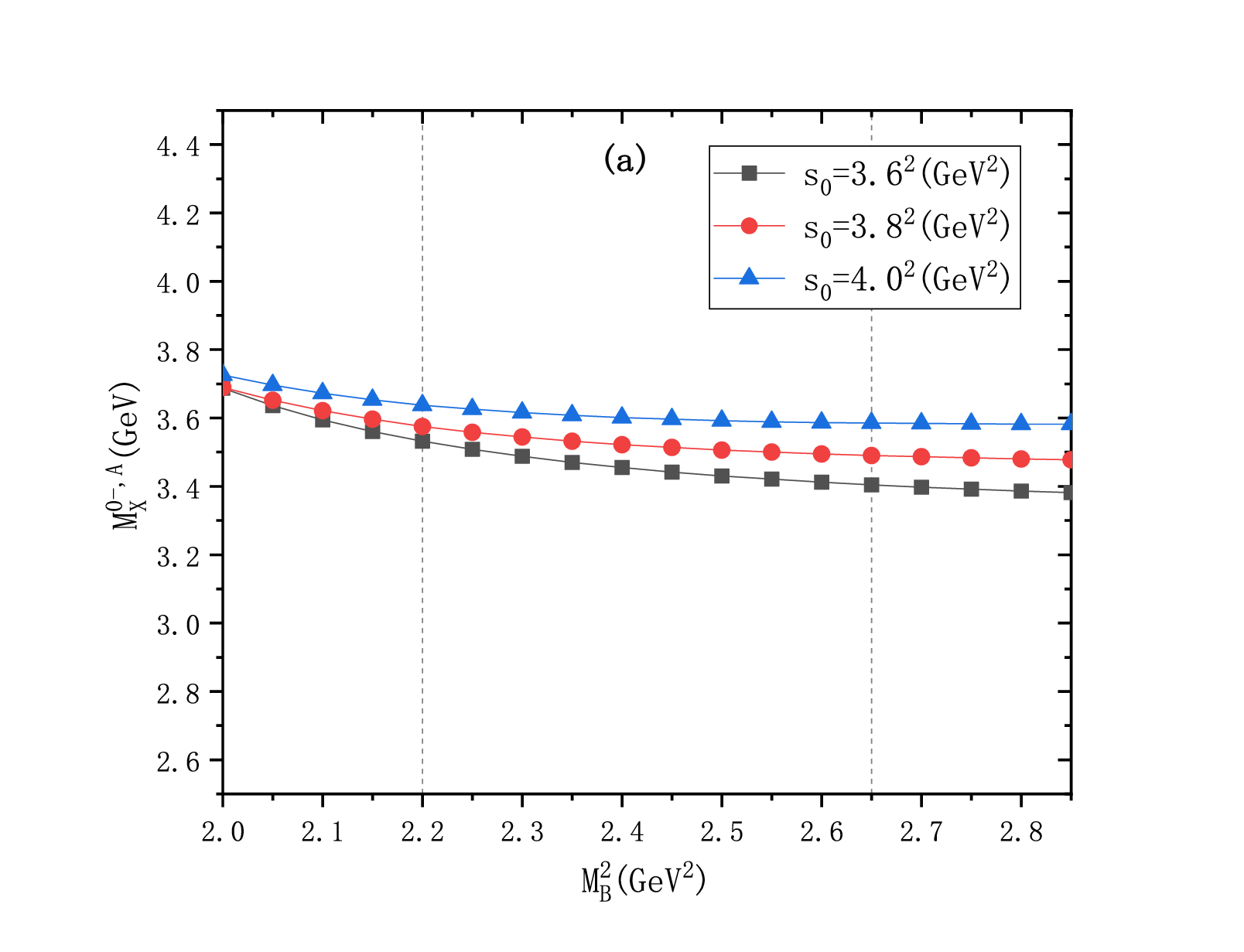}
\includegraphics[width=7.5cm]{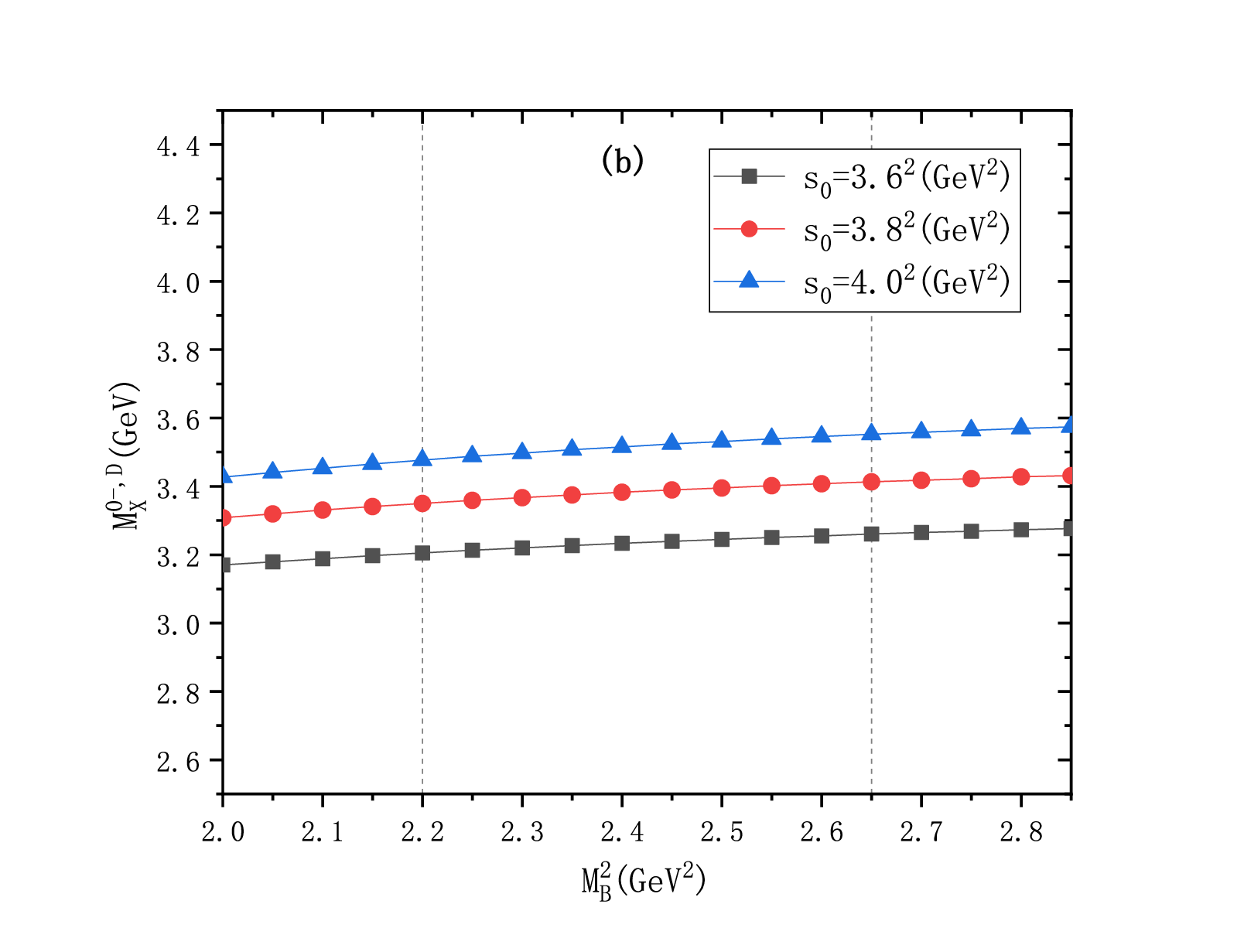}
\caption{~~~The same caption as in Fig.\ref{mass0+}, but for case $0^-$.}
\label{mass0-}
\end{center}
\end{figure}

\begin{figure}[!ht]
\begin{center}
\includegraphics[width=7.5cm]{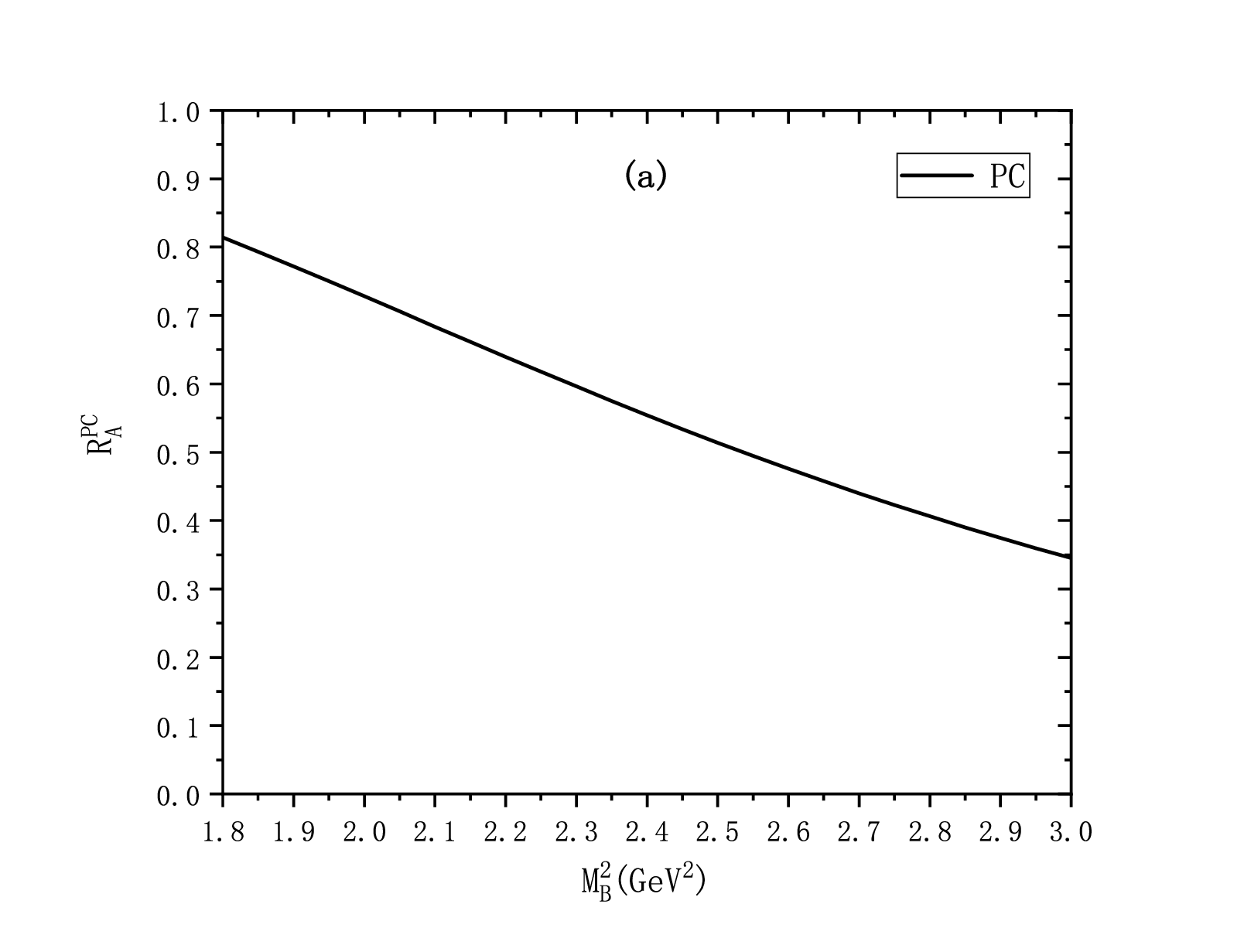}
\includegraphics[width=7.5cm]{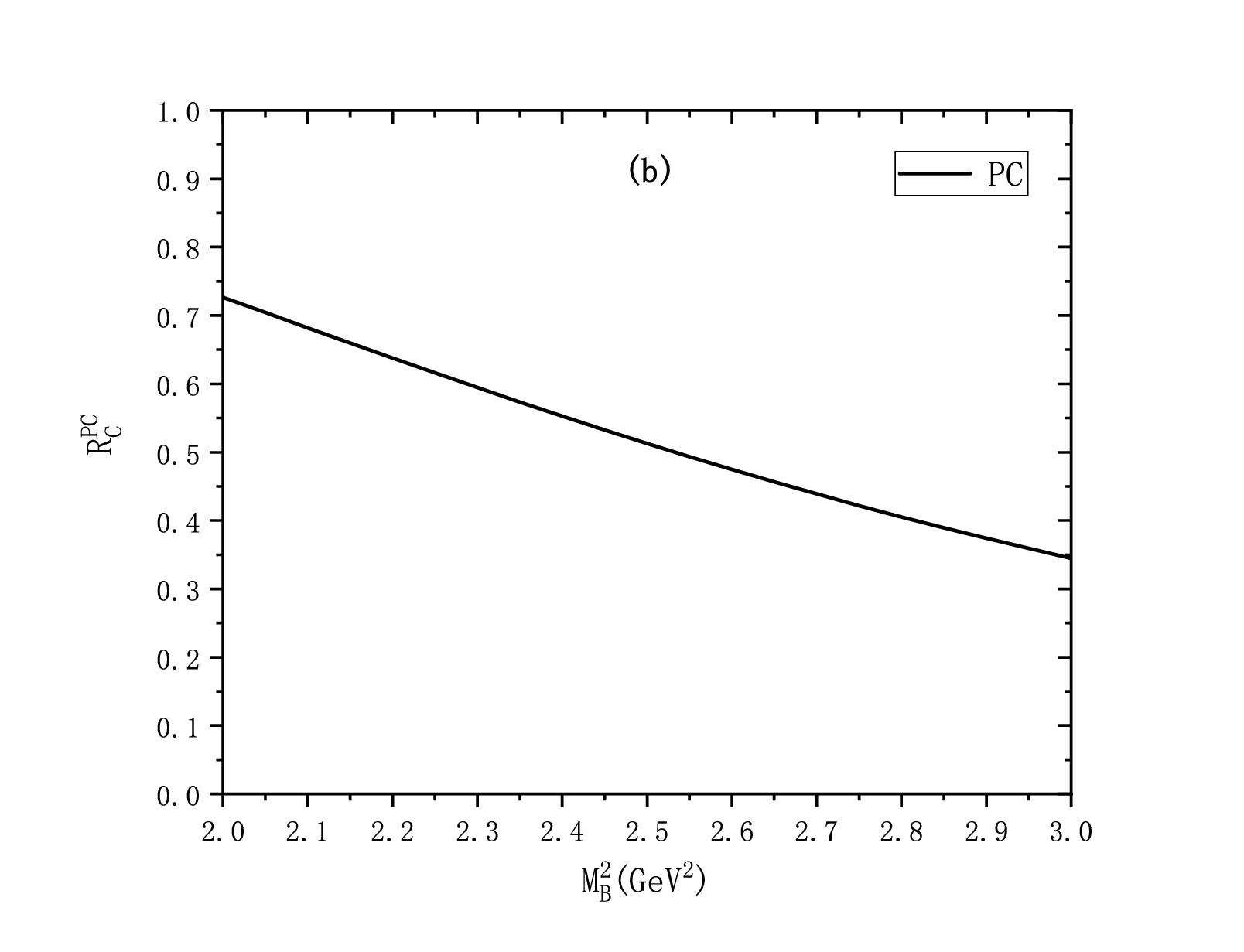}
\caption{~~~The same caption as in Fig.\ref{PC0+}, but for case $1^+$.}
\label{PC1+}
\end{center}
\end{figure}

\begin{figure}[!ht]
\begin{center}
\includegraphics[width=7.5cm]{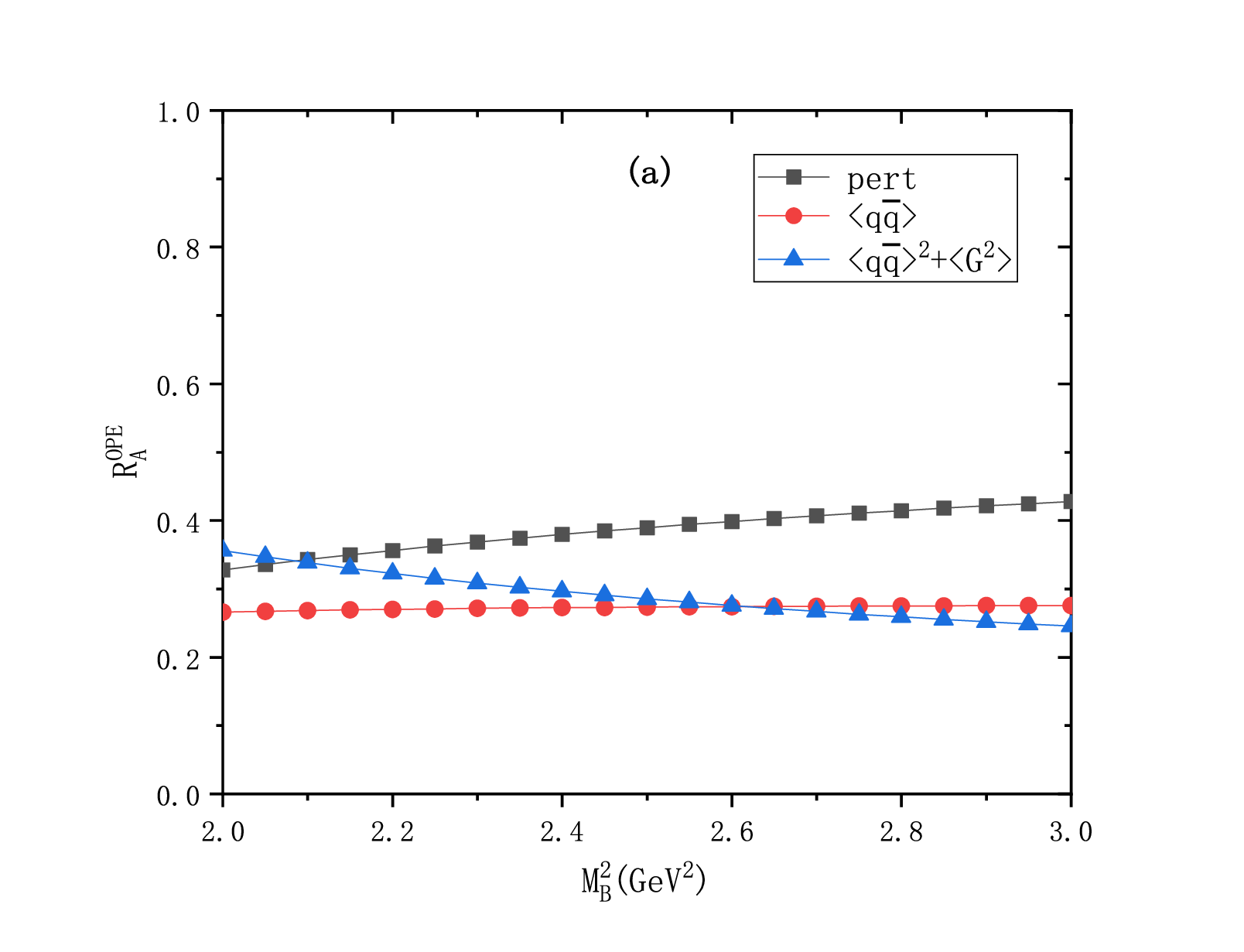}
\includegraphics[width=7.5cm]{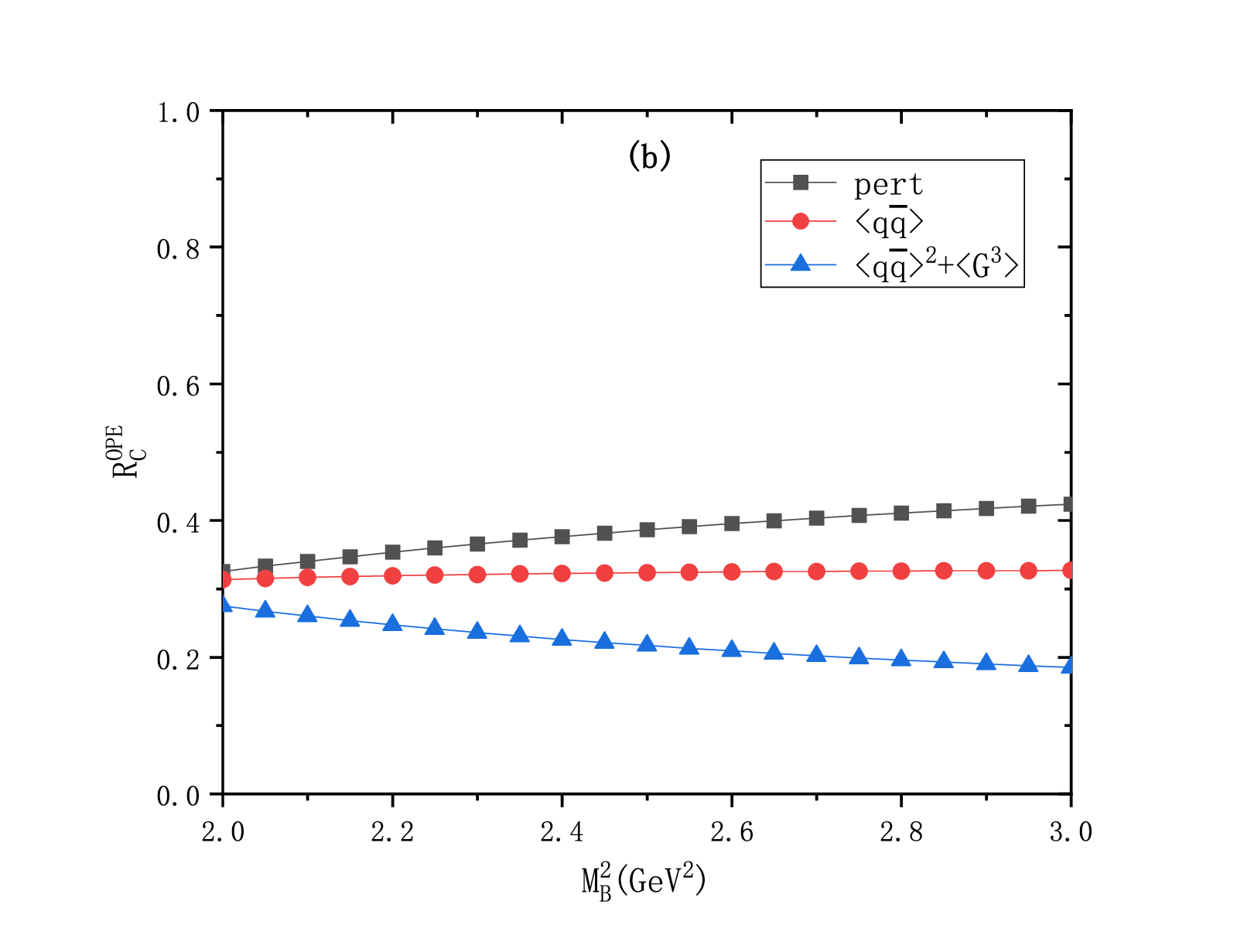}
\caption{~~~The same caption as in Fig.\ref{ROPE0+}, but for case $1^+$. }
\label{ROPE1+}
\end{center}
\end{figure}

\begin{figure}[!ht]
\begin{center}
\includegraphics[width=7.5cm]{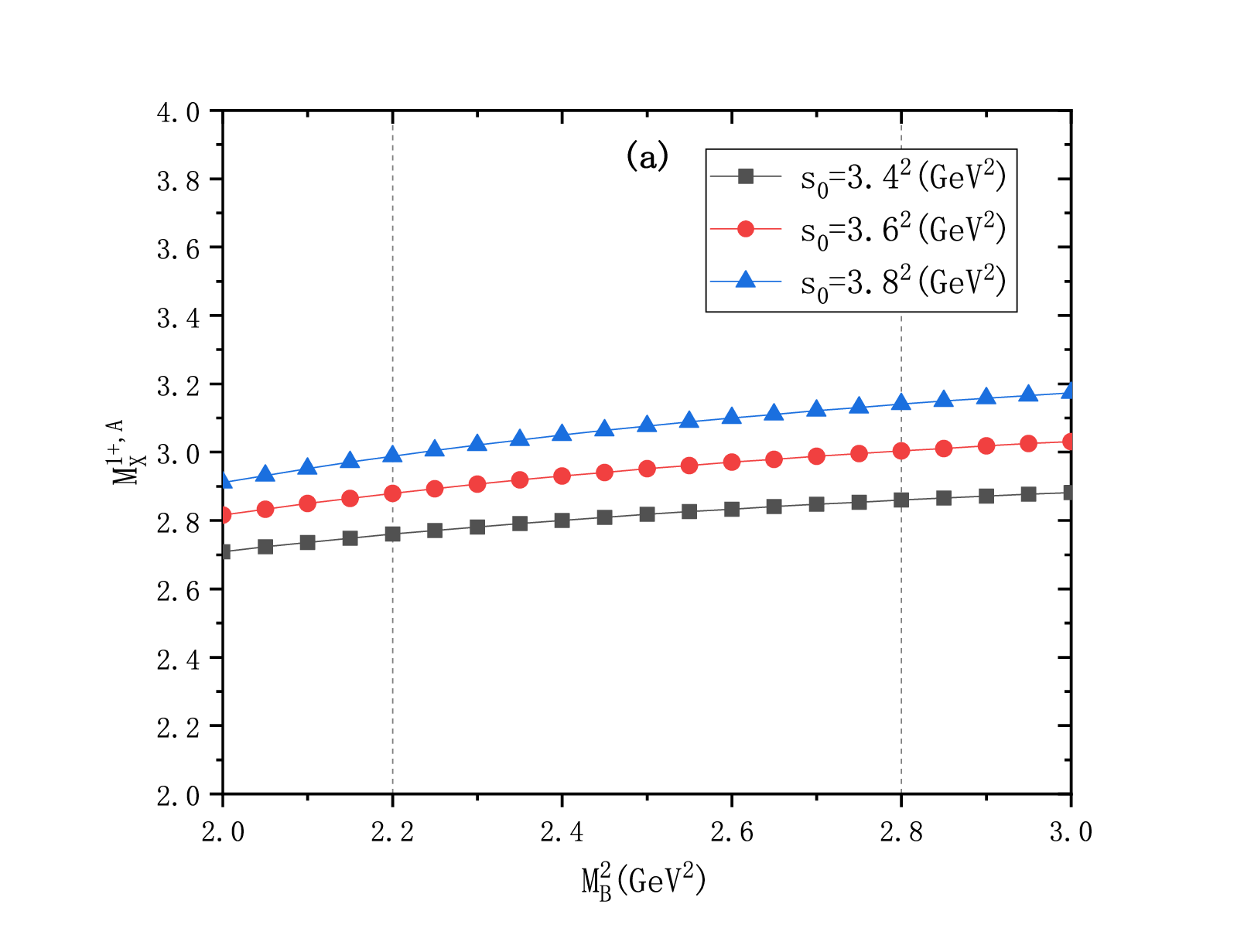}
\includegraphics[width=7.5cm]{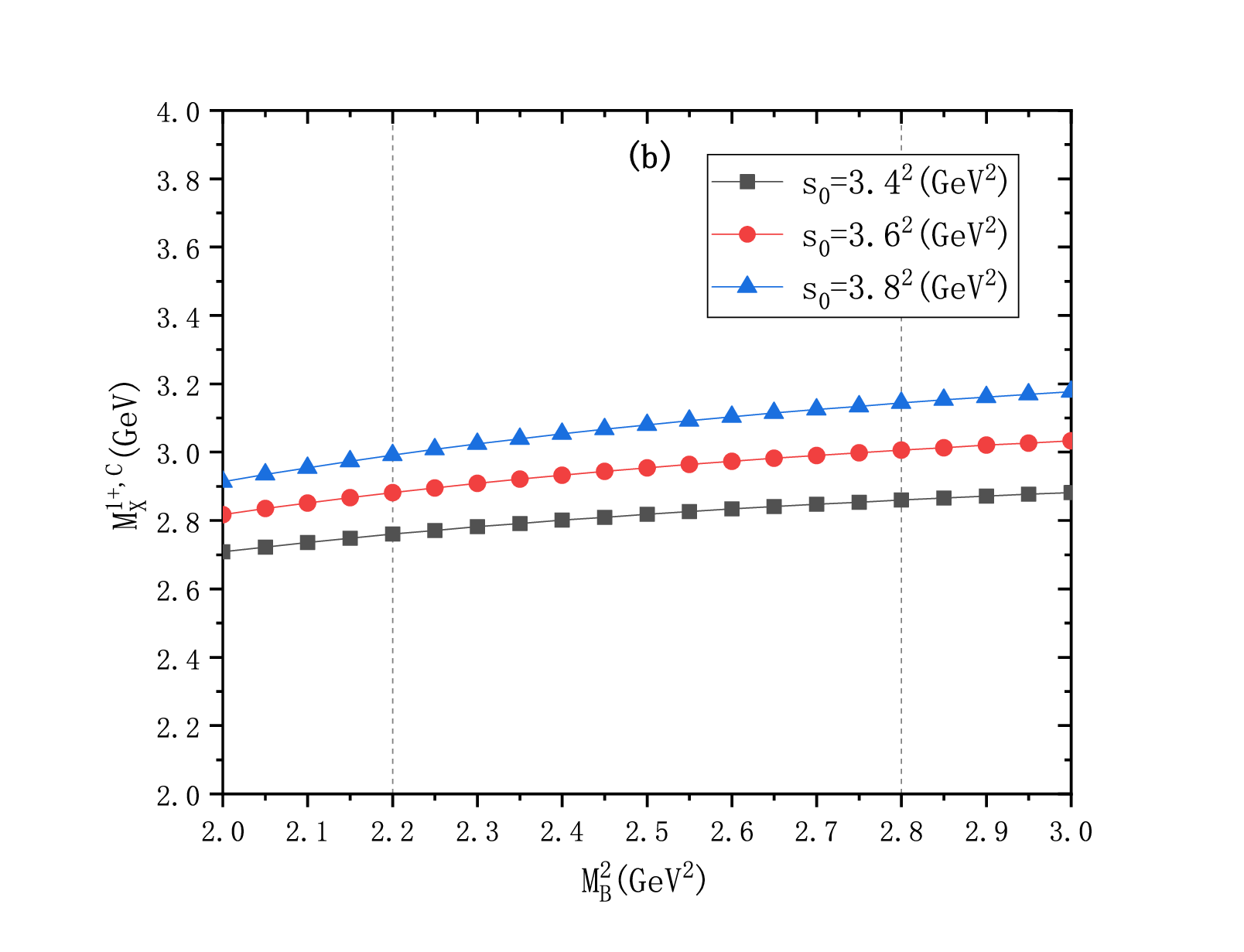}
\caption{~~~The same caption as in Fig.\ref{mass0+}, but for case $1^+$.} \label{mass1+}
\end{center}
\end{figure}

Using the same numerical analysis method, we can apply the above analysis process to cases with quantum numbers $0^-$, $1^+$ and $1^-$. We summarize the numerical results for all four cases in Table~\ref{tabss4}.
\begin{table}[h]
\centering
\caption{~~~Summary of the numerical results for tetraquark states in color octet-octet configuration, with quantum numbers $0^+$, $0^-$, $1^+$, and $1^-$.}
\begin{tabular}{ccccc}\hline
  $j^{J^{P}}_{X}$ & $M_B^2 \, (\rm{GeV}^2)$  & $\sqrt{s_{0}} \, (\rm{GeV})$ & PC & Mass $(\rm{GeV})$  \\ \hline
  $j^{0^{+}}_{A}$ & ${2.20\!-\!2.80}$ & {$3.6 \pm 0.2$} & ${(65\!-\!40)\%}$  & ${2.91_{-0.20}^{+0.20}}$ \\
  $j^{0^{+}}_{C}$ & ${2.00\!-\!2.60}$ & {$3.6 \pm 0.2$} & ${(66\!-\!40)\%}$  & ${2.98_{-0.21}^{+0.20}}$ \\
  $j^{0^{-}}_{A}$ & ${2.20\!-\!2.65}$ & {$3.8 \pm 0.2$} & ${(54\!-\!40)\%}$  & ${3.45_{-0.10}^{+0.12}}$ \\
  $j^{0^{-}}_{D}$ & ${2.35\!-\!2.00}$ & {$3.8 \pm 0.2$} & ${(54\!-\!40)\%}$  & ${3.20_{-0.19}^{+0.17}}$ \\
  $j^{1^{+}}_{A}$ & ${2.20\!-\!2.80}$ & {$3.6 \pm 0.2$} & ${(64\!-\!40)\%}$ & ${2.95_{-0.19}^{+0.19}}$ \\
  $j^{1^{+}}_{C}$ & ${2.20\!-\!2.80}$ & {$3.6 \pm 0.2$} & ${(64\!-\!40)\%}$ & ${2.95_{-0.19}^{+0.19}}$ \\
  $j^{1^{-}}_{A}$ & ${1.80\!-\!2.40}$ & {$3.6 \pm 0.2$} & ${(62\!-\!40)\%}$  & ${3.23_{-0.10}^{+0.11}}$ \\
  $j^{1^{-}}_{D}$ & ${1.85\!-\!2.45}$ & {$3.6 \pm 0.2$} & ${(63\!-\!40)\%}$  & ${3.23_{-0.11}^{+0.09}}$ \\ \hline
\end{tabular}
\label{tabss4}
\end{table}

\section{Discussion and conclusion}\label{sec:discussion}
In this work, we have investigated tetraquark states composed of $c \bar{s} u \bar{d}$ with a color octet-octet structure and quantum numbers $J^P=0^{\pm}$ and $1^{\pm}$ by applying the QCD sum rule method. For the $0^+$ case, we first constructed currents corresponding to the lowest orbital angular momentum states (i.e., S-wave) between color octets, resulting in four such currents as shown in currents (\ref{current-A})-(\ref{current-D}). Then, using the QCD sum rule method, we carried out the analytical calculation of the Operator Product Expansion (OPE) up to dimension 11 condensates, including contributions linear in the strange quark mass $m_s$. Finally, a numerical analysis was performed.

The numerical analysis indicates that currents (\ref{current-B}) and (\ref{current-D}) correspond to reasonable tetraquark states, with their masses being $M_X^A = 2.91_{-0.20}^{+0.20}~\text{GeV}$ and $M_X^C = 2.98_{-0.21}^{+0.20}~\text{GeV}$ respectively; currents (\ref{current-A}) and (\ref{current-C}) do not correspond to reasonable tetraquark states due to the lack of a reasonable parameter window.

By comparing with experimental results, our findings are consistent within the error range with the experimentally observed $T_{c\bar{s}0}^a(2900)^{0/++}$. Therefore, our calculations support classifying $T_{c\bar{s}0}^a(2900)^{0/++}$ as a tetraquark state with a color octet-octet structure.

Furthermore, based on the $J^P=0^+$ case, we also investigated tetraquark states made up of $c \bar{s} u \bar{d}$ with color octet-octet structure and quantum numbers $J^P=0^-$, $1^+$, and $1^-$. For each quantum number, we predicted two new hadron states awaiting experimental verification, with their masses as follows: $M_{0^{-}}^{A}=3.45_{-0.10}^{+0.12}~\text{GeV}$, $M_{0^{-}}^{D}=3.20_{-0.19}^{+0.17}~\text{GeV}$, $M_{1^{+}}^{A}=2.95_{-0.19}^{+0.19}~\text{GeV}$,
$M_{1^{+}}^{C}=2.95_{-0.19}^{+0.19}~\text{GeV}$, $M_{1^{-}}^{A}=3.23_{-0.10}^{+0.11}~\text{GeV}$ and $M_{1^{-}}^{D}=3.23_{-0.11}^{+0.09}~\text{GeV}$. With the advancement of experimental techniques and the accumulation of new data, these predicted results are expected to be confirmed in future experiments.

Our discovery not only provides theoretical support for the existence and internal structure of $T_{c\bar{s}0}^a(2900)^{0/++}$ but also highlights the QCD sum rules as a powerful tool for exploring the hadron spectrum of multi-quark states. This consistency also indicates the great potential of QCD sum rules in predicting the properties of exotic hadrons. This research and its conclusions further reinforce the significance of multi-quark states in particle physics, offering new directions for future experimental and theoretical studies.

\vspace{.7cm} {\bf Acknowledgments} \vspace{.3cm}

This work was supported by the Natural Science Foundation of Hebei Province under Grant No. A2023205038.


\newpage

\end{document}